\documentclass[aps,pre,reprint,superscriptaddress,letterpaper,twocolumn]{revtex4-1}

\usepackage{color}
\usepackage{amssymb}
\usepackage{amsmath}
\usepackage{bm}

\usepackage[ruled]{algorithm2e}
\usepackage{algorithmic}
\usepackage{setspace}

\usepackage{graphicx}

\usepackage{grffile}
\usepackage{subfigure}

\usepackage{hyperref}

\usepackage{cleveref}
\usepackage[utf8]{inputenc}
\usepackage{geometry}
\usepackage{float}

\begin{document}

\title{Inferring genetic fitness from genomic data}

\author{Hong-Li Zeng}
\email{hlzeng@njupt.edu.cn}
\affiliation{School of Science, and New Energy Technology Engineering Laboratory of Jiangsu Province, Nanjing University of Posts and Telecommunications, Nanjing 210023, China}
\affiliation{Nordita, Royal Institute of Technology, and Stockholm University, SE-10691 Stockholm, Sweden
}

\author{Erik Aurell}
\email{eaurell@kth.se}
\affiliation{KTH -- Royal Institute of Technology, AlbaNova University Center, SE-106 91 Stockholm, Sweden}%
\affiliation{
Faculty of Physics, Astronomy and Applied Computer Science, Jagiellonian University, 30-348 Krak\'ow, Poland
}

\begin{abstract}
The genetic composition of a naturally developing
population is considered as due to mutation, selection,
genetic drift and recombination.
Selection is modeled as single-locus terms (additive fitness)
and two-loci terms (pairwise epistatic fitness).
The problem is posed to infer epistatic fitness
from population-wide whole-genome data from a time series of a developing
population.
We generate such data in silico,
and show that in the Quasi-Linkage Equilibrium (QLE)
phase of Kimura, Neher and Shraiman, that pertains
at high enough recombination rates and low enough mutation rates,
epistatic fitness can be quantitatively correctly inferred using
inverse Ising/Potts methods.
\end{abstract}

\date{\today}

\maketitle

\section{Introduction}
\label{sec:introduction}
The last ten years has seen an explosion of interest in
results obtained from inferring terms in Ising or Potts
models from data~\cite{Nguyen-2017a}.
When applied in biological data analysis this approach
is known as
Direct Coupling Analysis (DCA)
and has led to a breakthrough in identifying
spatial contacts in proteins from protein sequence data~\cite{Weigt-2009a,Morcos-2011a,Stein-2015a,Michel-2017a,Cocco-2018a},
which in turn has been used to predict spatial contacts from the sequence data~\cite{Ovchinnikov-2017a,Michel-2017b,Ovchinnikov-2018a}.
DCA has also been used to identify
nucleotide-nucleotide contacts of RNAs~\cite{DeLeonardis-2015a},
multiple-scale protein-protein interactions~\cite{Gueudre-2016a,Uguzzoni-2017a},
amino acid--nucleotide interaction in RNA-protein complexes~\cite{Weinreb-2016a}
and synergistic effects not necessarily related to spatial contacts~\cite{Figliuzzi-2016a,Hopf-2017a,CouceE9026}.

Skwark \textit{et al} applied a version of DCA to
whole-genome sequencing data of a population of \textit{Streptoccoccus pneumoniae}~\cite{Skwark-2017a},
and were able to retrieve interactions
between members of the Penicillin-Binding Protein (PBP) family
of proteins. Schubert \textit{et al}
performed a similar analysis on data from \textit{Neisseria gonorrhoeae}~\cite{Schubert2019}.
Standard versions of DCA are rather compute-intensive for genome-scale inference tasks,
but methodological speed-ups~\cite{Puranen-2017a,GaoZhouAurell2018}
and alternative approaches~\cite{Pensar2019} have been developed.

It is thus computationally feasible to infer interactions between widely separated loci
from population-wide whole-genome data.
The question then arises if the results obtained are also biologically meaningful
\textit{i.e.} if, and when, they reflect underlying identifiable mechanisms.
In an earlier contribution~\cite{Gao2019}, one
of us argued that the reason should be sought in the
\textit{Quasi-Linkage Equilibrium} (QLE)
of Kimura~\cite{Kimura1956,Kimura1964,Kimura1965},
as later extended by Neher and Shraiman to statistical genetics on the genome scale~\cite{NeherShraiman2009, NeherShraiman2011}.
In this paper we test this explanation by simulating an evolving
population with known fitness parameters, treating the output of
the simulation as data, and then using DCA and the
Kimura-Neher-Shraiman theory (KNS) to infer fitness parameters.

We find that KNS indeed allows to infer fitness from data
in broad, but not global, parameter ranges.
Concerning the central parameter of KNS which is the
overall rate of recombination, the theory
works in an intermediate regime, while it fails at both a very high and very low rate.
We discuss these limitations as well as performance
when varying other parameters.
We also discuss implications for inferring epistatic interactions
from whole-genome population-wide data.

The paper is organized as follows.
In Section~\ref{sec:qle} we summarize relevant results of
the quasi-linkage equilibrium theory
and in Section~\ref{sec:model} we present the model
and the simulation methods.
In Section~\ref{sec:inference} we show tests of inference
procedure \eqref{KNS-eq2}.
Section~\ref{sec:phase-diagrams} synthesizes such tests to phase diagrams
of when fitness inference using \eqref{KNS-eq2} is possible/not possible
in these models, while
Section~\ref{sec:discussion} summarizes and discusses the results.
Supplementary technical details are given in three appendices.
In Appendix~\ref{a:FFPopSim} we give parameter settings for
simulations using the FFPopSim software introduced in Section~\ref{sec:model},
and in Appendices~\ref{a:nMF} and \ref{a:PLM} we give details on the
nMF and PLM inference procedures which we have used.

\section{A Quasi-Linkage Equilibrium primer}
\label{sec:qle}
The concept Quasi-Linkage Equilibrium (QLE) is built on the
distinction between Linkage Equilibrium (LE) and its opposite, Linkage Disequilibrium (LD).
For completeness we will first describe these concepts for the case of
two loci $A$ and $B$ where there can be, respectively, $n_A$ and $n_B$ alleles.
The configuration of one genome
with respect to $A$ and $B$
is then $(x_A,x_B)$ where $x_A$ takes values in $\{1,\ldots,n_A\}$
and $x_B$ takes values in $\{1,\ldots,n_B\}$.
The configuration of a population of $N$ individuals
is the set $\left[(x_A^{(s)},x_B^{(s)})\right]$ where $s$ ranges from $1$ to $N$.
This set defines the empirical probability distribution with respect to $A$ and $B$ as
\begin{equation}
P_{AB}(x_A,x_B) = \frac{1}{N}\sum_{s=1}^N \mathbf{1}_{x_A^{(s)},x_A} \mathbf{1}_{x_B^{(s)},x_B},
\end{equation}
where $\mathbf{1}_{a,b}$ is the Kronecker delta.
Similarly we can define $P_{A}(x_A) = \frac{1}{N}\sum_{s=1}^N \mathbf{1}_{x_A^{(s)},x_A}$,
and $P_{B}(x_B)$.
The distribution of genomes in a population over loci $A$ and $B$ is said to be in
\textit{Linkage Equilibrium} (LE) if the alleles $a_A$ and $x_B$ are independent under the
empirical distribution \textit{i.e.} if $P_{AB}(x_A,x_B)= P_{A}(x_A)P_{B}(x_B)$.
All other distributions are in \textit{Linkage Disequilibrium} (LD).

Independence implies that co-variances vanish.
That is, if $\left<\ldots\right>$
means averaging with respect to $P$, then in LE
\begin{equation}
C_{AB}(a,b) = \left<\mathbf{1}_{a,x_A} \mathbf{1}_{b,x_B} \right> - \left<\mathbf{1}_{a,x_A} \right>\left< \mathbf{1}_{b,x_B} \right> = 0
\end{equation}
The co-variance matrix $C_{AB}(a,b)$ always satisfies  $\sum_a C_{AB}(a,b)=\sum_b C_{AB}(a,b)=0$ and therefore has
$(n_A-1)(n_B-1)$ independent components.
For pairs of biallelic loci ($n_A=n_B=2$) it is convenient to label the alleles by
another set of variables $s$ that take values in $\pm 1$.
For this case $\left<\mathbf{1}_{1,x_A} \right>=\frac{1}{2}(1+\chi_A)$
and $\left<\mathbf{1}_{2,x_A} \right>=\frac{1}{2}(1-\chi_A)$
where $\chi_{A} = \left<s_A \right>$.
Similarly $C_{AB}(1,1)=-C_{AB}(1,2)=-C_{AB}(2,1)=C_{AB}(2,2)=\frac{1}{4}\chi_{AB}$ where
\begin{equation}
\chi_{AB} = \left<s_A s_B \right> - \left<s_A \right>\left< s_B \right>
\end{equation}
is the  co-variance between loci $A$ and $B$.
In LE all the coefficients $\chi_{AB}$ are zero.

\textit{Quasi-Linkage Equilibrium} is a subset of distributions in LD which are characterized by a distribution over genotypes of the form (for biallelic loci)
\begin{equation}
\label{eq:Ising}
P(\mathbf{s}) = \frac{1}{Z}\exp\left(\sum_i h_is_i + \sum_{i<j} J_{ij}s_is_j\right)
\end{equation}
where $Z$ (partition function) is the normalization.
Obviously \eqref{eq:Ising} is the Ising model of equilibrium statistical
mechanics, which like all maximum-entropy models is characterized by
its set of sufficient statistics that are here the
``magnetizations'' $\chi_{i}$ and ``correlations'' $\chi_{ij}$.
When all $\chi_{ij}$ (and all Ising parameters $J_{ij}$)
are small \eqref{eq:Ising} is close to the set of independent distributions
which characterize LE.

The fundamental insight of Motoo Kimura into this problem was that distributions
of the type \eqref{eq:Ising} appear naturally in population genetics
models which include biological recombination (or sex)~\cite{Kimura1956,Kimura1964,Kimura1965}.
If individual genomes are assimilated to configurations of a particle system,
a recombination event is akin to a collision. In a mechanism analogous to
relaxation to the equilibrium distribution in Boltzmann's equation,
\eqref{eq:Ising} then holds for a high enough rate of recombination.
The parameters $h_i$ and $J_{ij}$ of the genotype distribution
are hence in this theory \textit{consequences}
of a dynamical evolution law for the population
which takes into account recombination, mutations,
varying fitness and carrying capacity.
The mechanisms have been detailed previously~\cite{NeherShraiman2009,NeherShraiman2011,Gao2019} and will be reviewed again below in Section~\ref{sec:model}.

Fitness is in Kimura-Neher-Shraiman (KNS) theory a function
of the genotype
\begin{equation}
\label{eq:fitness}
F(\mathbf{s}) = \sum_{i} f_{i} s_i  + \sum_{i<j} f_{ij} s_i s_j
\end{equation}
where the linear coefficients $f_i$ are referred to
as \textit{additive contributions to fitness}
and the quadratic (pairwise) coefficients $f_{ij}$ are
\textit{epistatic contributions to fitness}.
The most important relation in KNS is
\begin{equation}
\label{KNS-eq1}
J_{ij} = \frac{f_{ij}}{rc_{ij}}
\end{equation}
where $J_{ij}$ are the parameters of the distribution in
\eqref{eq:Ising}, $f_{ij}$ are the model parameters in \eqref{eq:fitness},
$r$ is an overall rate of recombination,
and $c_{ij}$ is the probability that alleles at loci $i$ and $j$
are inherited from different parents.
When recombination is large this probability will be close to $\frac{1}{2}$
for most pairs of loci. Hence \eqref{KNS-eq1}
can be interpreted as a \textit{inference formula of epistatic fitness from genomic data}:
\begin{equation}
\label{KNS-eq2}
f^*_{ij} = J^*_{ij} \cdot rc_{ij}
\end{equation}
where $*$ indicates inferred value, $J^*_{ij}$ is determined from data, and the remaining
parameter $r$ is a proportionality.

Formula \eqref{KNS-eq1} is derived under the assumption that
mutation is low enough.
This is a potential confounder because if the mutation rate is strictly
zero the most fit genotype well eventually take over
in a finite population. When this has happened
one has instead of \eqref{eq:Ising} the much simpler result
\begin{equation}
\label{eq:frozen}
P(\mathbf{s}) = \mathbf{1}_{\mathbf{s},\hat{\mathbf{s}}}
\end{equation}
where $\hat{\mathbf{s}}$ is the most fit genotype.
Since there is no variability in data drawn from \eqref{eq:frozen}
there is therefore no way to infer parameters $J^*_{ij}$ from data, and
formula \eqref{KNS-eq2} cannot be used.

\section{Model and simulation methodology}
\label{sec:model}
We consider a population with a carrying capacity of $N$ individuals, which indicates the average size of the population is $N$.
Each individual characterized
by a genome of length $L$ ($L$ distinct loci).
For simplicity we assume as above biallelic loci, and encode the alleles as
$+1$ (major allele) and $-1$ (minor allele).
The genotype of an individual is then a string $\mathbf{s}=(s_1,\ldots,s_L)$
where each variable $s_i$ takes values in $\pm 1$.
An evolving population is described by a
time-dependent normalized probability
distribution $P\left(\mathbf{s},t\right)$.
This section describes the mechanisms
by which these changes
are assumed to occur~\cite{NeherShraiman2009,NeherShraiman2011,Gao2019},
and how we simulate those changes on the population level
using the FFPopSim software package~\cite{FFPopSim}.

\paragraph{Mutations} are random changes of the $s_i$, assumed to occur independently
at different loci and in different individuals.
We further assume that they are characterized by one overall rate $\mu$,
which is the probability that any one allele at any locus in any individual
changes per unit time (per generation).
They hence lead to a simple gain-loss master equation
\begin{equation}
\label{eq:dP-mutations}
\frac{dP}{dt}|_{mut} = \mu\sum_{i=1}^L P(F_i\mathbf{s})-P(\mathbf{s})
\end{equation}
where $F_i$ (flip operator on locus $i$) acts on strings as
\begin{equation}
\label{eq:flip}
F_i(s_1,\ldots,s_i,\ldots,s_L)=(s_1,\ldots,-s_i,\ldots,s_L)
\end{equation}

\paragraph{Fitness}
Fitness variations were introduced above in \eqref{eq:fitness}.
The effect of fitness is that the higher the value $F\left(\mathbf{s}\right)$
for an individual, the higher is
the number of expected offspring of this individual
in the next generation.
On the population level this gives
\begin{equation}
\label{eq:dP-fitness}
\frac{dP}{dt}|_{fit} = P(\mathbf{s})\left(F(\mathbf{s})-\left<F\right>\right)
\end{equation}
where the average fitness, with respect to the
given distribution, is
\begin{equation}
\label{eq:F-average}
\left< F \right>=\sum_{\mathbf{s}}P(\mathbf{s})F(\mathbf{s})
\end{equation}
The coefficients in \eqref{eq:fitness} are hence \emph{rates} in this model, with dimension inverse time.

We characterize the epistatic contributions
to fitness as a model parameter by
\begin{equation}
\label{eq:sigma}
\sigma(\mathbf{f}) = \sqrt{\frac{2}{L(L-1)}\sum_{i<j} f^2_{ij}}
\end{equation}
where we have assumed that the average of the $f_{ij}$ is zero.

We note that in~\cite{NeherShraiman2009,NeherShraiman2011} fitness variability
was characterized by the standard deviation
of the fitness function $F\left(\mathbf{s}\right)$ with respect to $P$.
Although this better reflects the fluctuations of fitness in a population
it is a derived parameter (it depends on $P$) and hence less convenient
in numerical testing.

\paragraph{Genetic drift} is the randomness from a finite number of individuals in
the population, where some genotypes may propagate and multiply from one generation to the next,
and others die out.

In our set-up genetic drift can be
formulated together with fitness
by first having every individual $a$ give rise
to a random number $k_a$ of offspring (identical copies of themselves).
These numbers $k_a$ are Poisson distributed with rates $e^{\Delta t F_a}$
where $\Delta t$ is the (short) generation time and $F_a$ is the fitness
of individual $a$.
The total number of individuals is then
$N'= \sum_{a=1}^N k_a$ which is brought back to $N$
by either copying a further $N-N'$ uniformly randomly chosen
individuals in the new population, or by randomly eliminating $N'-N$ individuals.
Mutations can be incorporated in the same frame-work by
first randomly flipping each allele in every individual with
probability $\mu \Delta t$.

The interplay between mutations and genetic drift are encoded in
Fisher-Wright models, and has been studied since the beginning of population genetics~\cite{Fisher-book},
and has been reviewed many times, \textit{e.g.} in~\cite{Blythe2007}.
On the level of distributions genetic drift gives rise to
diffusion terms (Kimura's diffusion approximation).

\paragraph{Recombination} is a way
for two individuals (parents) pool their genomes to give
rise to an individual with a genome that is a combination of the parents.
The model presented in~\cite{NeherShraiman2009,NeherShraiman2011}
applies to haploid yeast.
In such organisms an individual is ordinarily
described by
one genome sequence $\mathbf{s}$ (haploid phase).
At the time of mating each individual additionally
makes a second copy of their genome (``mating body'')
so that they temporarily hold two identical copies of $\mathbf{s}$
(diploid phase).
In recombination two mating bodies merge to make
one new individual which carries one genome sequence $\mathbf{s}$
containing a mix of the genetic
information from the parents.
By this process the total number of individuals grows by
the number of pairs that have mated which is balanced
by randomly eliminating a fraction of the new population so that the total remains $N$.

As discussed in~\cite{Gao2019} the model in~\cite{NeherShraiman2009,NeherShraiman2011}
basically also applies to forms of bacterial recombination.
In that case two individuals (two bacteria)
with genotypes $\mathbf{s}_1$ and $\mathbf{s}_2$
recombine
with a rate $r Q(\mathbf{s}_1,\mathbf{s}_2)$ where $r$ an overall factor
and $Q(\mathbf{s}_1,\mathbf{s}_2)$ a relative rate.
The outcome of the recombination is
two individuals (two bacteria) $\mathbf{s}_1'$ and $\mathbf{s}_2'$
which can be specified by the
indicator variable $\mathbf{\xi}$:
\begin{eqnarray}
  \mathbf{s}_1':\quad   s^{(1)'}_i &=&  \xi_i s_i^{(1)}  + (1-\xi_i) s_i^{(2)} \\
  \mathbf{s}_2':\quad   s^{(2)'}_i &=& (1-\xi_i) s_i^{(1)} + \xi_i s_i^{(2)}
\end{eqnarray}
and this outcome of the recombination happens with probability $C(\mathbf{\xi})$.
The total rate of the individual event producing
$\mathbf{s}_1'$ and $\mathbf{s}_2'$
from $\mathbf{s}_1$ and $\mathbf{s}_2$
is hence $r Q(\mathbf{s}_1,\mathbf{s}_2)C(\mathbf{\xi})$,
and the change of the distribution over genotypes due to recombination is
\begin{eqnarray}
 \label{eq:Potts-recomb-bact}
  \frac{d}{dt} P(\mathbf{s})|_{rec}
=  r \sum_{\mathbf{\xi},\mathbf{s}'}   C(\mathbf{\xi})
  \big[&Q&{{(\mathbf{s}_1,\mathbf{s}_2) P(\mathbf{s}_1) P(\mathbf{s}_2)  }} \nonumber \\
 &-& {{Q(\mathbf{s},\mathbf{s}') P(\mathbf{s})P(\mathbf{s}') }}\big]
\end{eqnarray}
From a physical point of view this type of
recombination is analogous to a collision process
where two-genome distributions on the right hand side of
\eqref{eq:Potts-recomb-bact} have been assumed to factorize.

A central quantity in the theory is
the probability that in the offspring of two individuals
that have recombined the alleles on two loci $i$
and $j$ have been inherited from different parents:
\begin{equation}
  \label{eq:cij}
c_{ij} = \sum_{\mathbf{\xi}} C(\mathbf{\xi})\left(\xi_i(1-\xi_j)+(1-\xi_i)\xi_j\right)
\end{equation}
If there is an \textit{even} number of recombination events between
$i$ and $j$ then the alleles at these two loci are inherited from
\textit{the same} parent. If on the other hand
there is an \textit{odd} number of recombination events between
$i$ and $j$ then the alleles at these two loci are inherited from
\textit{different} parents.
Therefore we also have
\begin{equation}
\label{eq:cij-2}
c_{ij} = p(1;i,j)+p(3;i,j)+p(5;i,j)+\ldots
\end{equation}
where $p(k;i,j)$ is the probability to have $k$
recombination events between $i$ and $j$.
These probabilities will in general depend on $i$ and $j$ and the
distribution $C(\mathbf{\xi})$ in a nontrivial manner.

A simple assumption that can be evaluated easily is
if recombination can happen anywhere on a genome
between two neighboring loci with uniform probability $\rho$.
The number of recombination events between $i$ and $j$
is then binominally distributed
and
\begin{eqnarray}
\label{eq:cij-3}
c_{ij} &=& \frac{1}{2}\left(1-\left(1-2\rho)^{|i-j|}\right)\right) \nonumber \\
&\approx& \frac{1}{2}\left(1-e^{-2\rho|i-j|}\right)
\end{eqnarray}
where the second line holds if $\rho|i-j|$ is order unity and $|i-j|$ is large.
For pairs of loci sufficiently far apart
we have $c_{ij} \approx \frac{1}{2}$, and (\ref{eq:cij-3}) is the approximation we will use in the simulation.

\paragraph{Parameters}
The three important parameters we chose to study are $\mu$, $\sigma$ and $r$
representing mutations, selection and recombination respectively.

Naturally this choice is stylized (or simplified), as modifications and additions
can be made to all three mechanisms.
Mutation rates in real organisms can and will vary between loci,
and will typically not be the same in both directions.
Gaussian distributed fitness variance are well characterized by the standard deviation,
but other model distributions would depend on more parameters, for an example
see \textit{e.g.} the ``random power-law distribution'' introduced in~\cite{GaoZhouAurell2018}.
As discussed above
recombination for closely enough spaced loci depends even in the simplest
model on parameter $\rho$. More generally
recombination can have the
same overall rate, but still differ greatly
in \textit{e.g.} the lengths of genomic sub-sequences interchanged between two individuals
and along the genomes (``recombination hotspots'').

\paragraph{FFPopSim}
For simulations we have used the FFPopSim software package~\cite{FFPopSim} with settings
as described in Appendix~\ref{a:FFPopSim}.
FFPopSim allows for different types of recombination, each
of which is described by a set of parameters; these parameters are given in Appendix~\ref{a:FFPopSim}.

\paragraph{Mutations, carrying capacity and length of simulation}
As discussed above around \eqref{eq:frozen} in the absence of
mutations the most fit genotype will eventually take over
a finite population. When this has happened all variability is
lost, and it is no longer possible to infer epistatic contributions
to fitness.

\begin{figure}[htpb]
\centering
\subfigure[$\mu=0.0005$]{
\begin{minipage}[t]{0.31\linewidth}
\centering
\includegraphics[width=\textwidth]{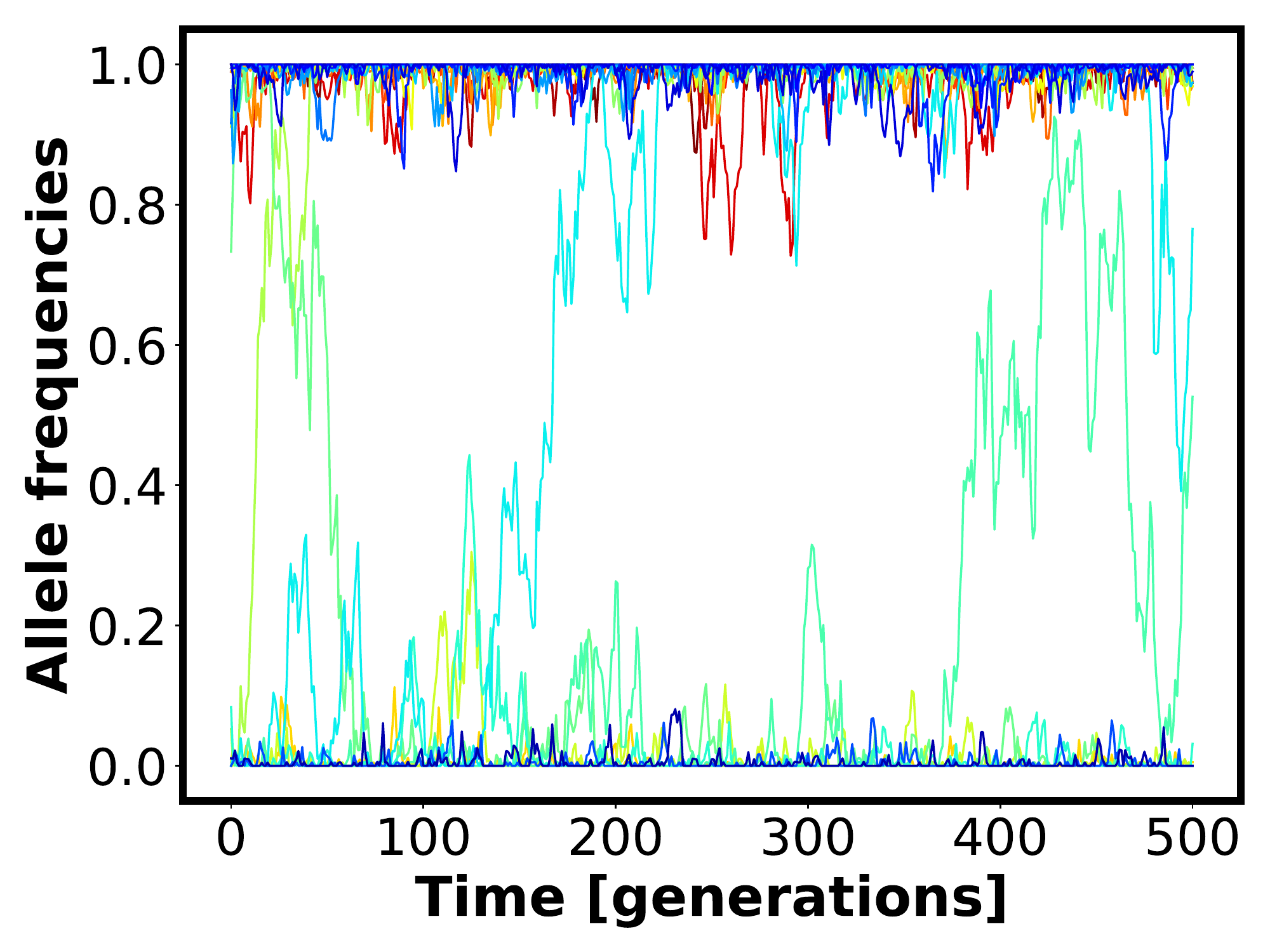}
\end{minipage}%
}
\subfigure[$\mu=0.005$]{
\begin{minipage}[t]{0.31\linewidth}
\centering
\includegraphics[width=\textwidth]{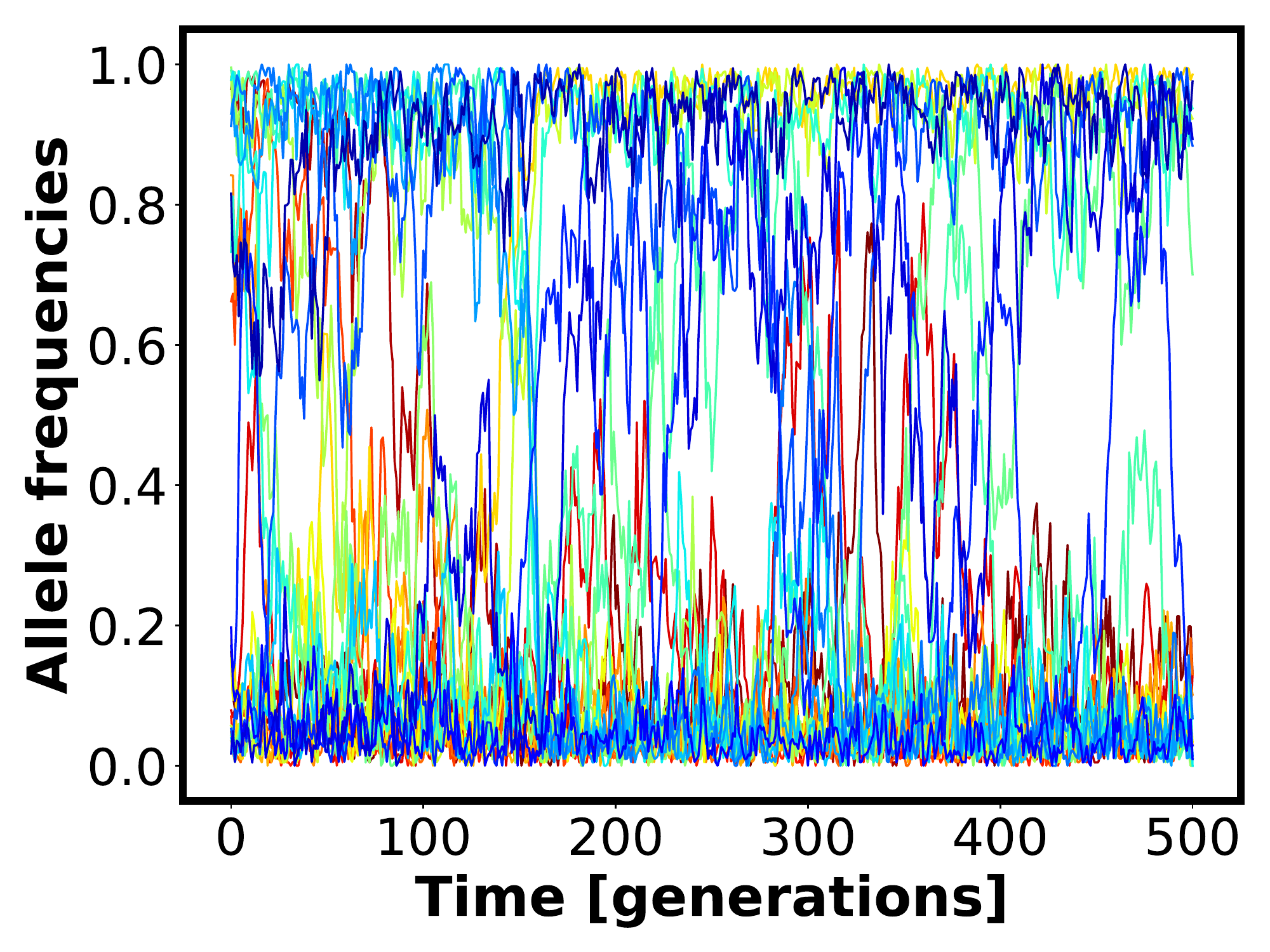}
\end{minipage}%
}
\subfigure[$\mu=0.3$]{
\begin{minipage}[t]{0.31\linewidth}
\centering
\includegraphics[width=\textwidth]{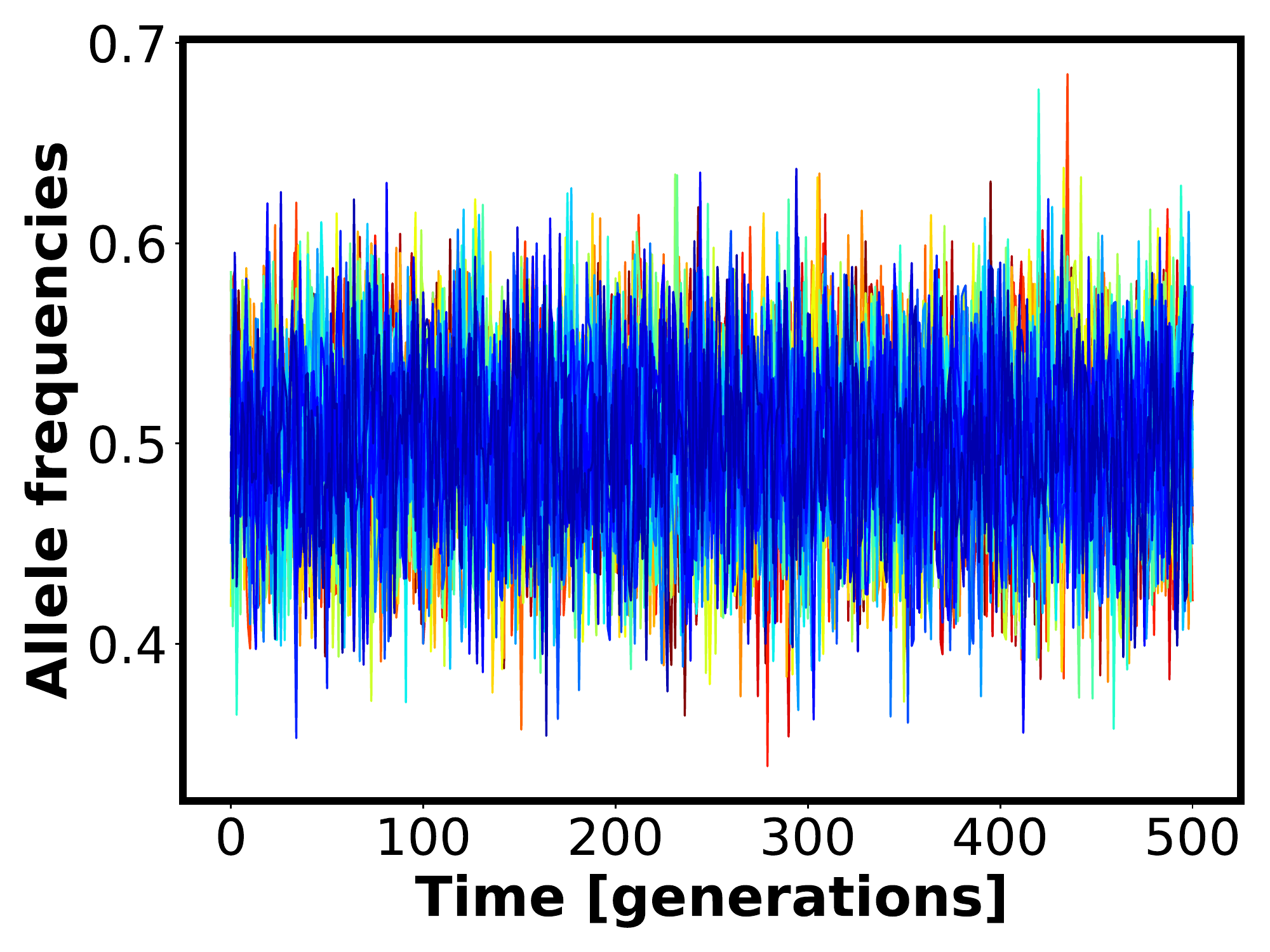}
\end{minipage}%
}
\centering
\caption{Frequency of the first allele $f_i[1]$ for all loci as function of time at different mutation rates $\mu$. Data output every five generations.
This quantity is related to $\chi_i$ defined in text by
$\chi_i=f_i[1]-1$. Population size $N=200$, number of loci $L=25$
, recombination rate $r=0.05$, cross-over rate $\rho=0.5$ and fitness variability $\sigma=0.002$.
Pairwise epistatic fitness parameters $f_{ij}$ are distributed as Gaussians (Sherrington-Kirkpatrick model), additive fitness parameters $f_i$ are zero.
The number of generations $T = 5 \times 500$.
Panel (a): mutation rate $\mu = 0.0005$. For most loci the distribution
is frozen to one value (0 or 1) for most of the simulation time; longer simulations would have been needed to gather enough data for inference.
Panel (b): mutation rate $\mu = 0.005$. For many loci the distribution
fluctuates away from the limiting value, although for most loci
frequencies do not change over in the available simulation time.
Panel (c): mutation rate $\mu = 0.3$. For all loci the
frequencies change over multiple times.
}\label{fig:time-series}
\end{figure}

When mutation rates are non-zero but small the distribution should most of the time
also resemble \eqref{eq:frozen}, and only for long
enough times will the population at some loci swift between the available alleles.
Fig.~\ref{fig:time-series}a shows that for sufficiently low mutation rates
the population this is indeed the case.
For higher mutation rates (Fig.~\ref{fig:time-series}b)
frequencies fluctuate faster between the extremes,
and at sufficiently high mutation rates (Fig.~\ref{fig:time-series}c)
frequencies for most of the time hover around the entropically dominant
configurations, where approximately half are up and half are down.

The size of the carrying capacity $N$, the mutation rate $\mu$ and the simulation time (generation) $T$
hence exert a combined effect on the simulations
which impacts on tests of KNS theory.
When $N$ is very large, as it is likely to always be in real data from bacterial or fungal
populations, one will
have sufficient variability to estimate ensemble locus-locus correlations
from data from one time. In such a case $T$ can be ignored.

At intermediate values of $N$, as one will typically have in a simulation,
then at moderate mutation rates,
Fig.~\ref{fig:time-series}a,
there is not enough variability in the population to estimate
locus-locus correlations from data at one time.
It is only if one uses data from different times that one
can meaningfully test KNS theory in this range.
$T$ is hence here a meaningful parameter relative to
$N$ and $\mu$. For example,
in Fig.~\ref{fig:time-series}b, there may be enough information
in the time series to estimate correlations (and fitness)
but in Fig.~\ref{fig:time-series}a there clearly cannot be since there is very little variation.

At high enough mutation rate, Fig.~\ref{fig:time-series}c
there will again be enough variability to estimate correlations
from data from one time, and $T$ can again be ignored.
However, the quantitative aspects of
KNS theory such as inference formula
\eqref{KNS-eq2} have been derived under the
assumption that mutation is a weaker effect than recombination.
Therefore in this range KNS cannot be expected to be
quantitatively correct, whether applied to data from one time, or
to data from a time series.

\section{Inference methods and their performance in different phases}
\label{sec:inference}
The inverse Ising problem is to infer model parameters
in the distribution \eqref{eq:Ising} from data drawn independently from that same distribution.
A large collective effort reviewed and summarized
in~\cite{Nguyen-2017a} has led to a detailed understanding of when such
a procedure can (or can't) work.
The dimensions of the problem are
\begin{itemize}
\item number of samples ($N$)
\item the average size and distribution of the underlying parameters ($h$ and $J$)
\item the inference procedure used
\end{itemize}
The main inference procedures in use
are ``naive'' mean-field (nMF), pseudo-likelihood maximization (PLM) and
Boltzmann machines. The latter is an iterative
method to find the maximum likelihood estimate of the parameters.
In general performance and computational cost increases along this list.
Boltzmann machines in particular are not feasible for large enough instances
due to the exponential complexity of the computation of ensemble averages.
We will therefore not consider that method further here.

As discussed above, snap-shot data at one time from a simulation of an evolving population
at finite $N$ is different from data drawn from an Ising/Potts distribution.
Even if carrying capacity $N$ is large enough to estimate correlations,
these same correlations will fluctuate in time with amplitude $N^{-\frac{1}{2}}$,
see~\cite{NeherShraiman2011} (Section VI and Appendix D).
For these reasons we will use variants of nMF and PLM that use
all the data in a simulation, a distinction which we underline by the qualifier \textit{alltime-}.

nMF means to treat the model as Gaussian and is hence given by the matrix inversion formula
\begin{equation}
\label{eq:nMF}
 J^*_{ij} = -(\chi)^{-1}_{ij}
\end{equation}
where $\chi_{ij} = \langle s_i s_j \rangle - \chi_i \chi_j $, and $\chi_i= \langle s_i \rangle$ are correlations and means respectively.
\begin{figure}[h!]
\centering
\subfigure[$\mu=0.0005$]{
\begin{minipage}[t]{0.31\linewidth}
\centering
\includegraphics[width=\textwidth]{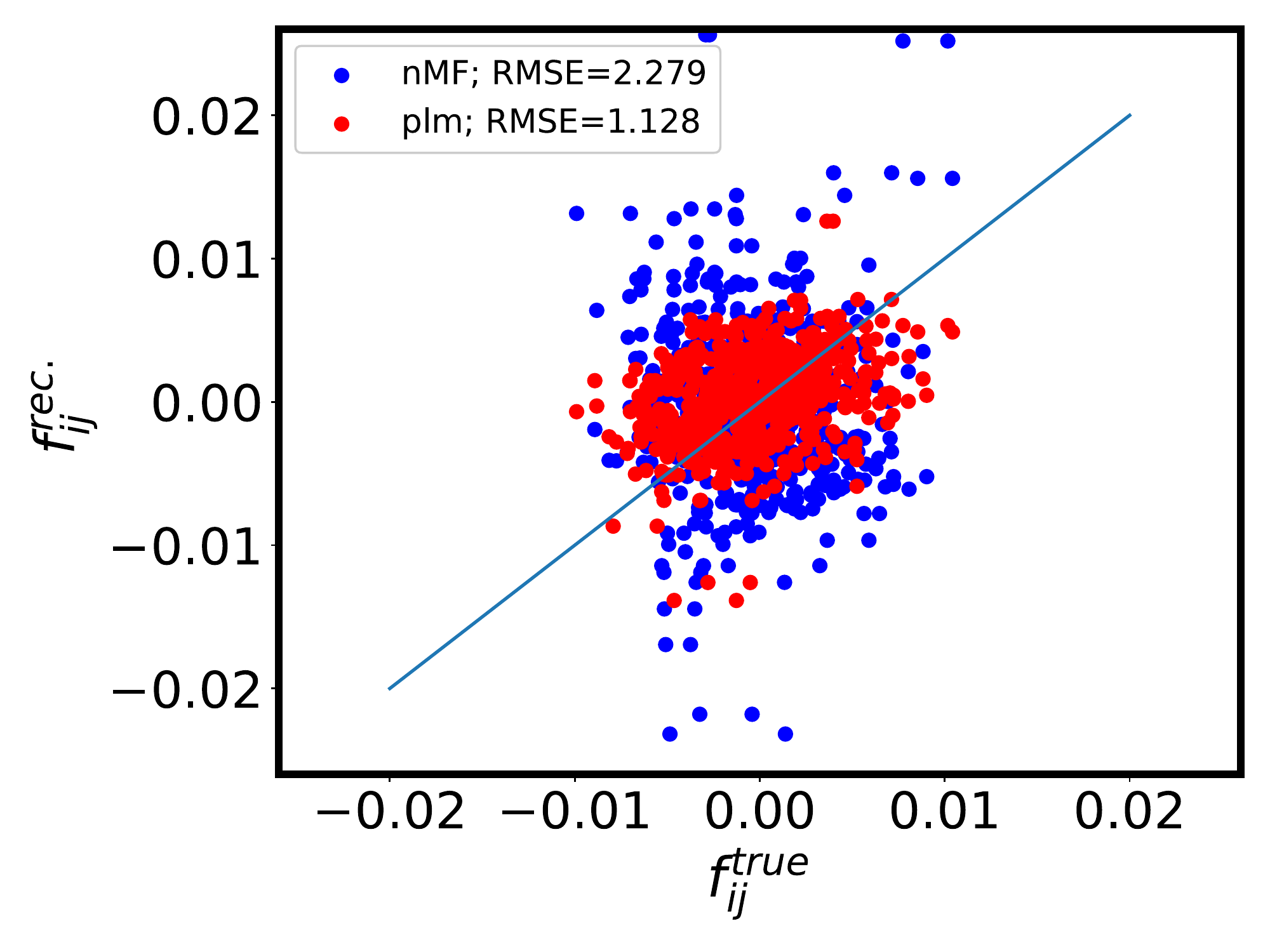}
\end{minipage}%
}
\subfigure[$\mu=0.005$]{
\begin{minipage}[t]{0.31\linewidth}
\centering
\includegraphics[width=\textwidth]{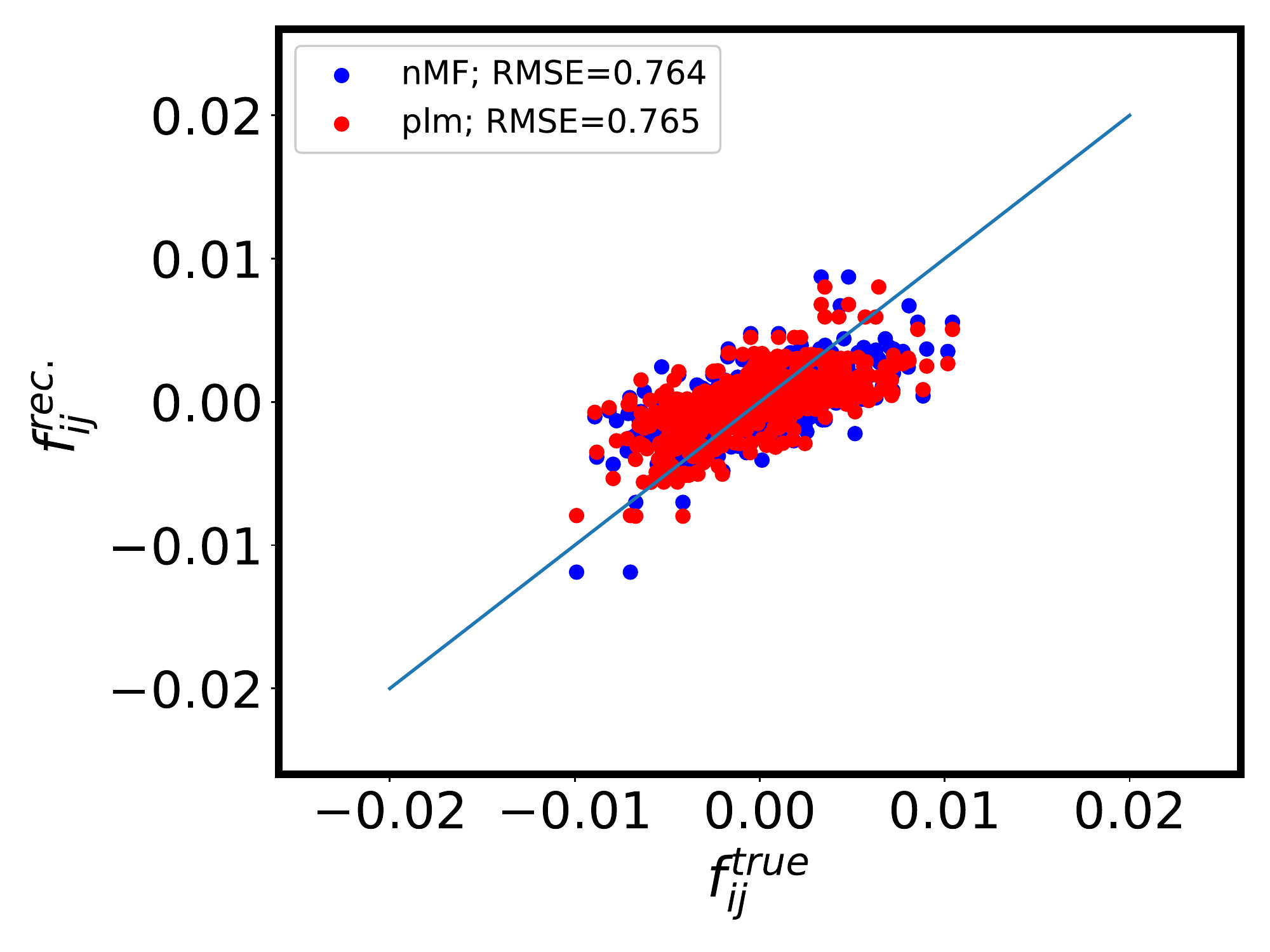}
\end{minipage}%
}
\subfigure[$\mu=0.3$]{
\begin{minipage}[t]{0.31\linewidth}
\centering
\includegraphics[width=\textwidth]{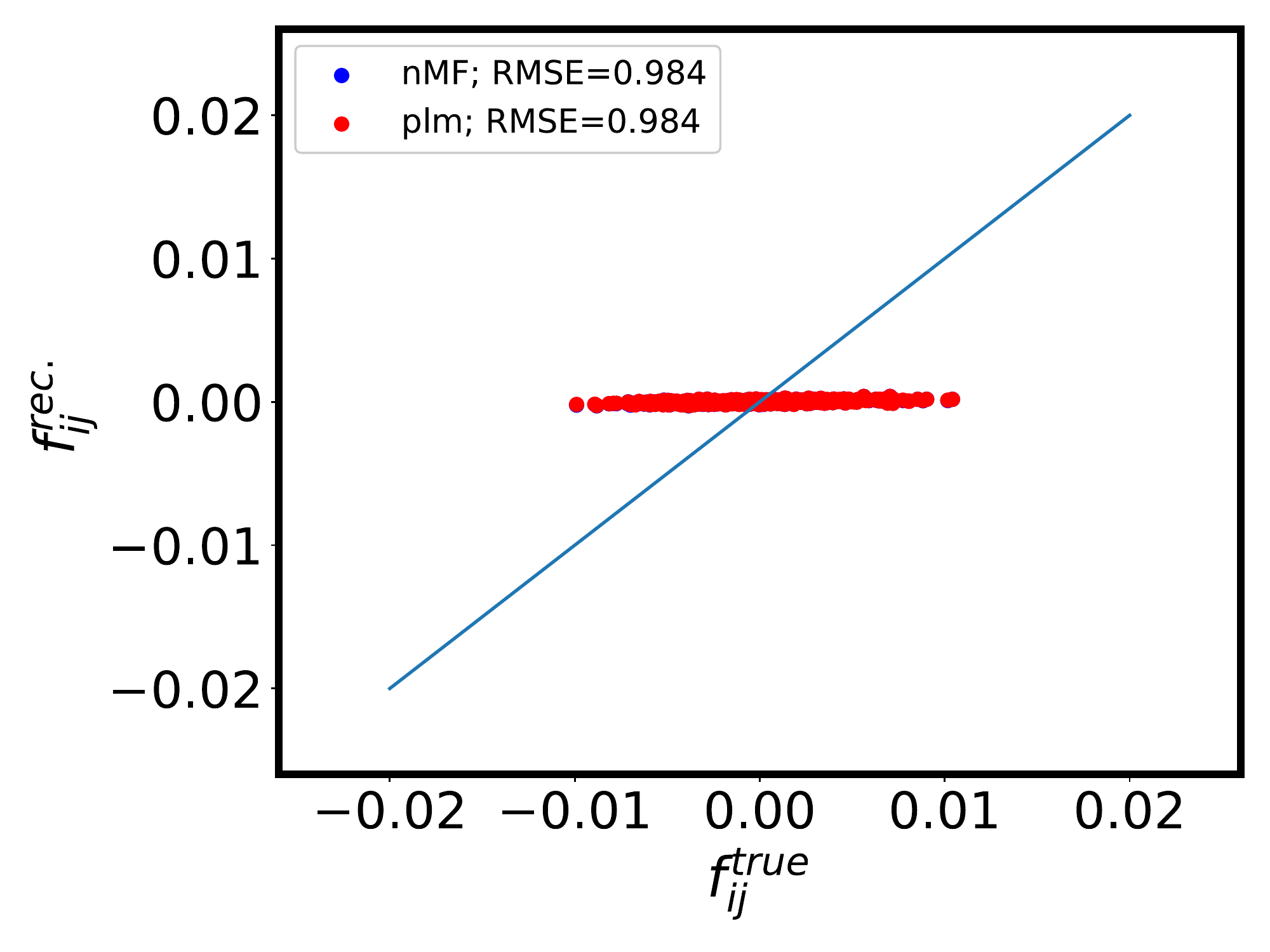}
\end{minipage}%
}
\centering
\caption{Scatter-plots of inferred fitness parameters $f^*_{ij}$ vs.
model fitness parameters $f_{ij}$, all parameter values the same as in Fig.~\protect\ref{fig:time-series}.
Inference of Ising parameters $J^*_{ij}$ is by the \textit{alltime-nMF} and \textit{alltime-PLM}
procedures (described in main text).
Inferred fitness parameters $f^*_{ij}$ are determined from
Ising parameters $J^*_{ij}$ by the Kimura-Neher-Shraiman formula, \protect\eqref{KNS-eq2}.
Panel (a): at low mutation rates the inferred fitness values are unrelated to
the underlying model parameters (``a cloud of points'').
Panel (b): at intermediate mutation rates there is sufficient variability in the data
and the procedure is fairly accurate.
Panel (c): at high mutation rates the inferred fitness values are everywhere small
such that plot is an almost horizontal line.
}\label{fig:scatter-plots}
\end{figure}

\textit{Alltime-nMF} means that we pool all the individuals from the whole population
and all times into one large data set. From this we compute $\chi_{ij}$ and $\chi_i$,
and from there we infer the parameters  $J^*_{ij}$ using \eqref{eq:nMF}.
Compared to nMF on data from one time, the computational cost of \textit{alltime-nMF}
scales linearly in $T$ (time needed to compute the averages and correlation functions).
Further details on \textit{alltime-nMF} are given in Appendix~\ref{a:nMF}.
We note that one can also consider the opposite approach of
first using \eqref{eq:nMF} on the data from each time, and
then averaging the inferred $J^*_{ij}$ over time. For parameter ranges where
\textit{alltime-nMF} works as an inference procedure we find that this second approach gives similar results (data not shown).


In contrast to nMF, PLM is a method to estimate parameters from conditional probabilities of one data item $s_i$ (one spin)
conditioned on all the others ($\mathbf{s}_{\setminus i}$). For the Ising model this conditional probability is
\begin{equation}
\label{eq:Ising-conditional}
P(s_i|\mathbf{s}_{\setminus i}) = \frac{e^{h_is_i + \sum_{j\neq i} J_{ij}s_is_j}}{\sum_{u}e^{h_iu + \sum_{j\neq i} J_{ij}us_j}},\end{equation}
where $u=\pm 1$ is the possible states of $s_i$.
Compared to the full probability \eqref{eq:Ising} $P(s_i|\mathbf{s}_{\setminus i})$ only depends on a much smaller
set of parameters, and is normalized in a way that is much simpler to deal with.
Given a number of samples, assumed independent, one can then maximize the corresponding log-likelihood function
\begin{equation}\label{eq:Ising-PLM}
\begin{aligned}
  PL_i \left(h_i, \{J_{ij} \} \right)
&= h_i\left< s_i\right> + \sum_{j\neq i} J_{ij} \left< s_is_j\right>\\
 &- \left< \log \sum_{u}e^{h_i u + \sum_{j\neq i} J_{ij}us_j}\right>
\end{aligned}
\end{equation}

Maximizing $PL_i$ will give inferred parameter ``as seen from $i$'', symbolically written $h_i^{*i}$ and $J_{ij}^{*i}$. Since in fact there
is only one Ising model parameter $J_{ij}$, PLM must be complemented by an output procedure for which a standard choice is
\begin{equation}
\label{eq:PLM}
J^*_{ij} = \frac{1}{2}\left(J^{*i}_{ij} + J^{*j}_{ij} \right)
\end{equation}
\textit{Alltime-PLM} means that we consider the whole population
at all times in the simulation as $N\cdot T$ samples from the same probability distribution, and use those to
compute the log-likelihood functions $PL_i$ in $\eqref{eq:Ising-PLM}$.
The computational cost of \textit{alltime-PLM} is considerably heavier than \textit{alltime-nMF}
essentially because there are many terms in the pseudo-log partition function, the last term in \eqref{eq:Ising-PLM}.
In Appendix~\ref{a:PLM} we give further details on \textit{alltime-PLM}, including estimates of computation times.

Using the same parameters and data as in Fig.~\ref{fig:time-series},
one then finds that inference of fitness is not possible at low
mutation rates: the scatter-plot in Fig.~\ref{fig:scatter-plots}a
is but a cloud of points with no visible trend.
At intermediate mutation rates Fig.~\ref{fig:scatter-plots}b
inference by KNS \protect\eqref{KNS-eq2} works,
while for large mutation rates Fig.~\ref{fig:scatter-plots}c
\protect\eqref{KNS-eq2} does not work again.

\section{Phase diagrams}
\label{sec:phase-diagrams}
Large-scale tests of inference as in Fig.~\ref{fig:scatter-plots}
requires a quantitative criterion for when inference is successful or not.
Here we will use the \textit{normalized $L_2$ distance} given by
\begin{equation}
\label{eq:normalized-L2}
\epsilon = \sqrt{\frac{\sum_{i<j}\left(f^*_{ij}-f_{ij}\right)^2}{\sum_{i<j} f_{ij}^2}} .
\end{equation}
When $\epsilon$ is much smaller than one then inference is successful and the scatter-plots will look
more or less like Fig.~\ref{fig:scatter-plots}b (or better).
On the other hand $\epsilon$ can take values of around one or larger, either because $f^*_{ij}$ and $f_{ij}$ are about the same
size but uncorrelated, as in Fig.~\ref{fig:scatter-plots}a, or if $f^*_{ij}$
appears a a function of $f_{ij}$ but of another form.
This is so in in Fig.~\ref{fig:scatter-plots}c where $f^*_{ij}$
is close to zero for all pairs of loci, but other dependencies yielding similar $\epsilon$ could also be present.

\begin{figure}[htbp]
\centering
\subfigure[nMF]{
\begin{minipage}[t]{0.48\linewidth}
\centering
\includegraphics[width=\textwidth]{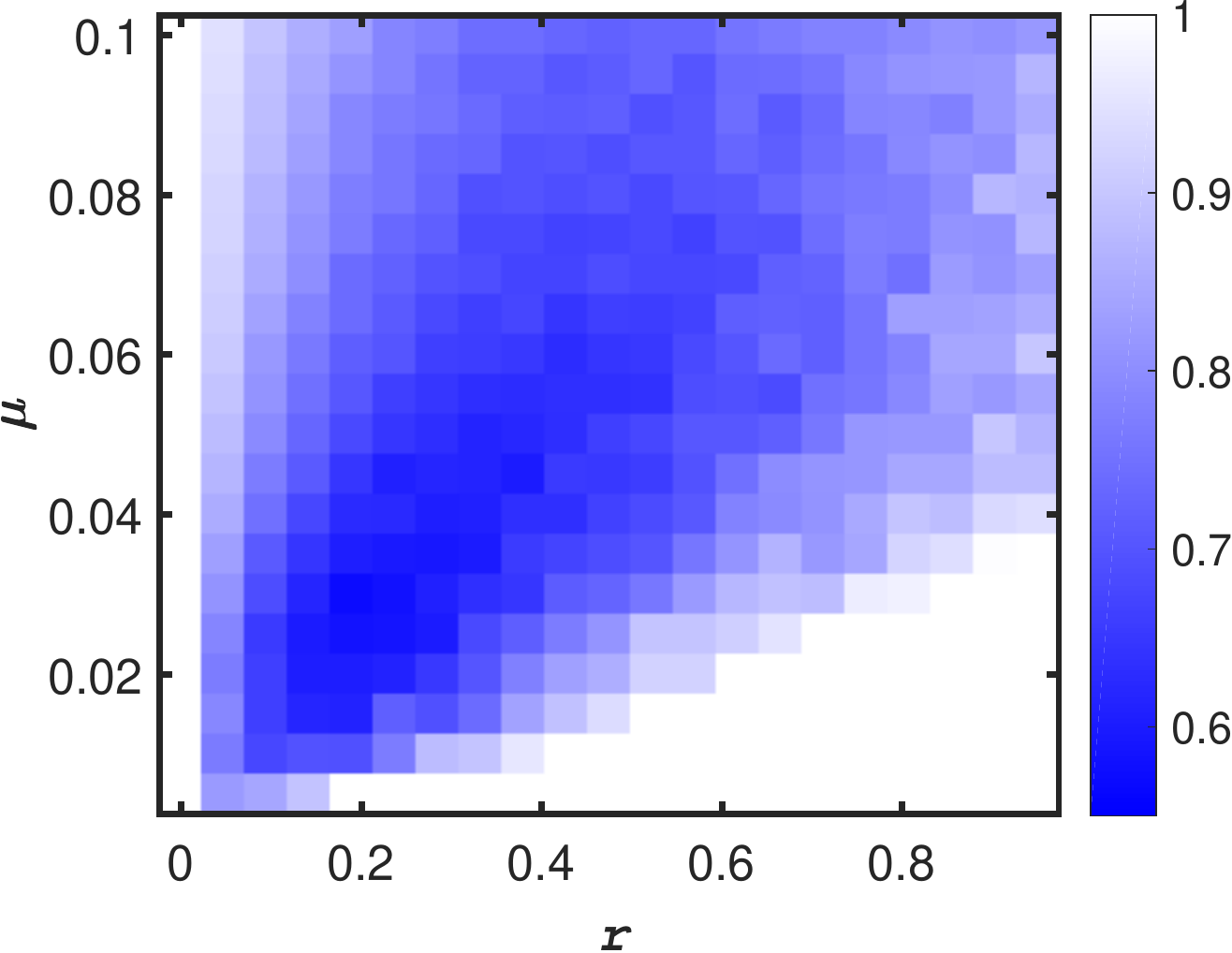}
\end{minipage}%
}
\subfigure[PLM]{
\begin{minipage}[t]{0.48\linewidth}
\centering
\includegraphics[width=\textwidth]{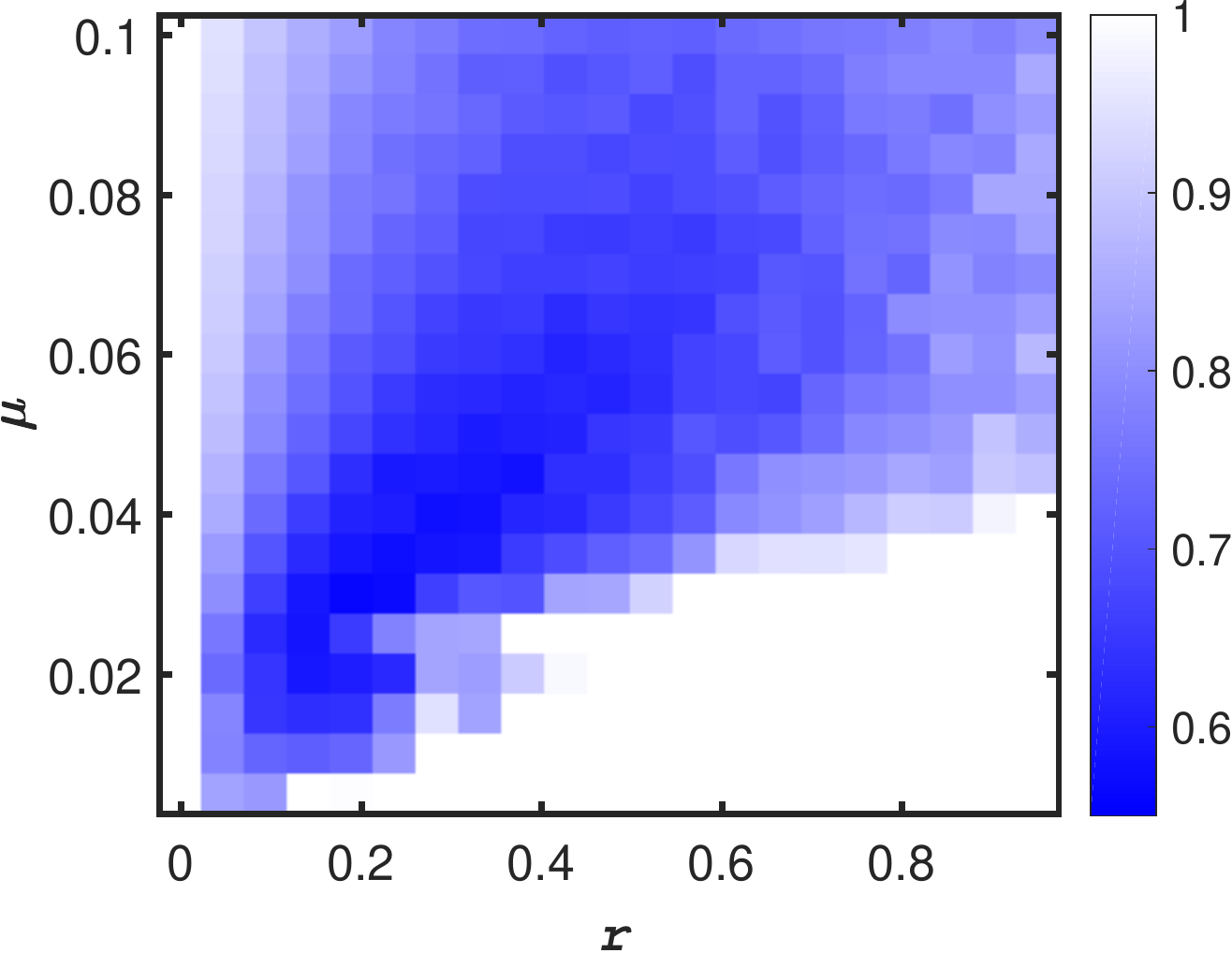}
\end{minipage}%
}
\caption{Phase diagram of epistatic fitness inference quality color coded by the reconstruction error $\epsilon$ in (\protect\ref{eq:normalized-L2}).
Panel (a): fitness estimated by formula \protect\eqref{KNS-eq2} with \emph{alltime-nMF} procedure.
Panel (b): fitness estimated by formula \protect\eqref{KNS-eq2} with \emph{alltime-PLM} procedure.
By the color coding both schemes work in a very broad range of parameters. For large recombination
inference does not work for the reasons illustrated in Fig.~\ref{fig:different-rec.-rate}c, while
for small recombination the assumptions underlying \protect\eqref{KNS-eq2} are not satisfied, as shown in Fig.~\ref{fig:different-rec.-rate}a.
}\label{fig:phase-diagram-mutation-vs-recombination}
\end{figure}

We can now display the phase diagram by plotting $\epsilon$ in \eqref{eq:normalized-L2}
as function of parameters, color coded by the value of $\epsilon$.
Fig.~\ref{fig:phase-diagram-mutation-vs-recombination} shows this
for the two inference formulas in the plane of mutation rate $\mu$ versus recombination rate $r$.
With \eqref{KNS-eq2}  we qualitatively expect that inference will not work for sufficiently low recombination, because then this assumptions behind this formula are not satisfied (as will be shown below in Fig.~\ref{fig:different-rec.-rate}a) and also not for very high recombination while low mutation neither, as will be illustrated in Fig.~\ref{fig:different-rec.-rate}c.
Essentially this turns out to be correct.
In Fig.~\ref{fig:phase-diagram-mutation-vs-recombination} fitness variations are everywhere small compared to mutations and recombination.

\begin{figure}[htbp]
\centering
\subfigure[$r=0.05$]{
\begin{minipage}[t]{0.3\linewidth}
\centering
\includegraphics[width=\textwidth]{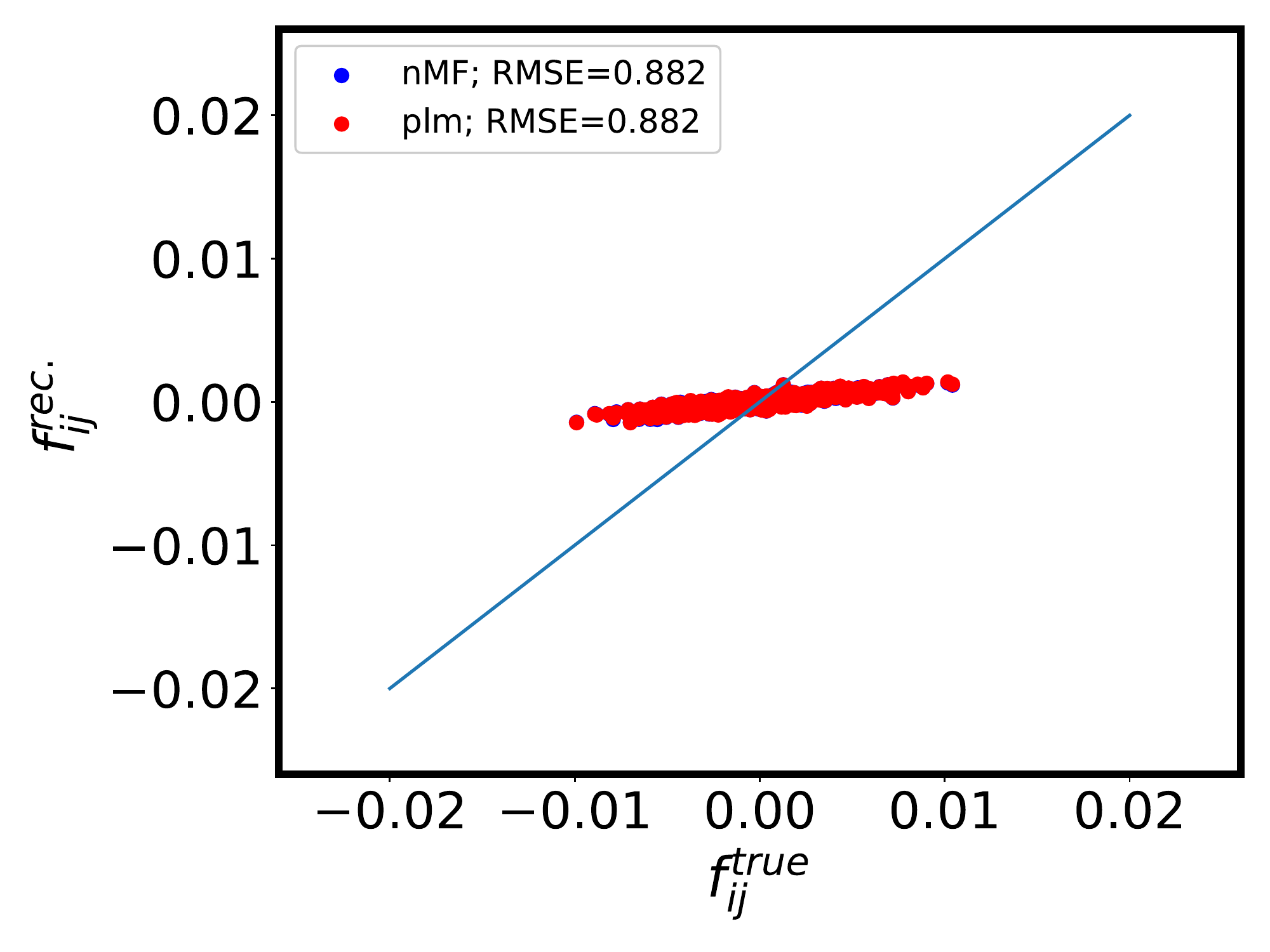}
\end{minipage}%
}
\subfigure[$r =0.5$]{
\begin{minipage}[t]{0.3\linewidth}
\centering
\includegraphics[width=\textwidth]{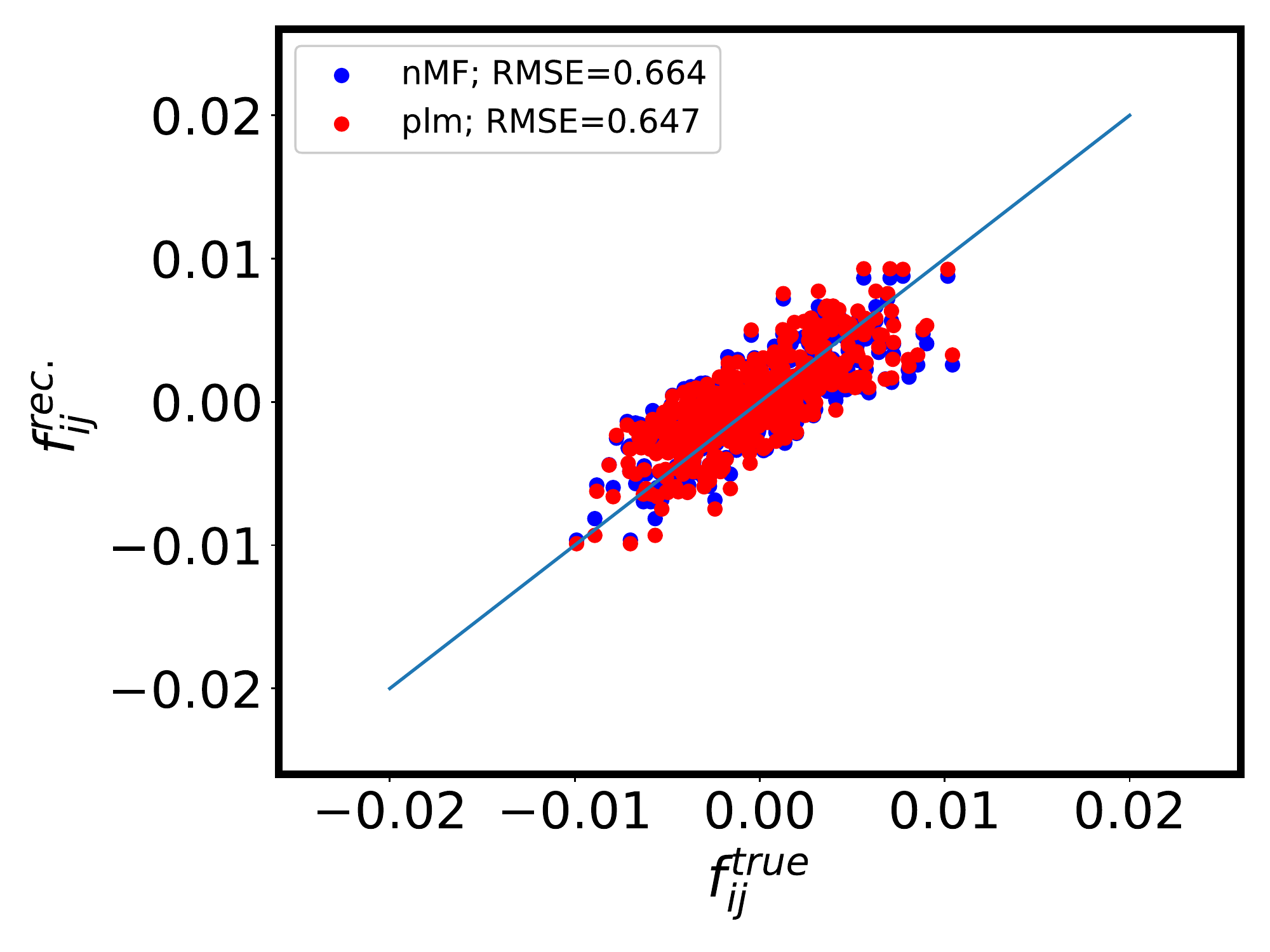}
\end{minipage}%
}
\subfigure[$r =1$]{
\begin{minipage}[t]{0.3\linewidth}
\centering
\includegraphics[width=\textwidth]{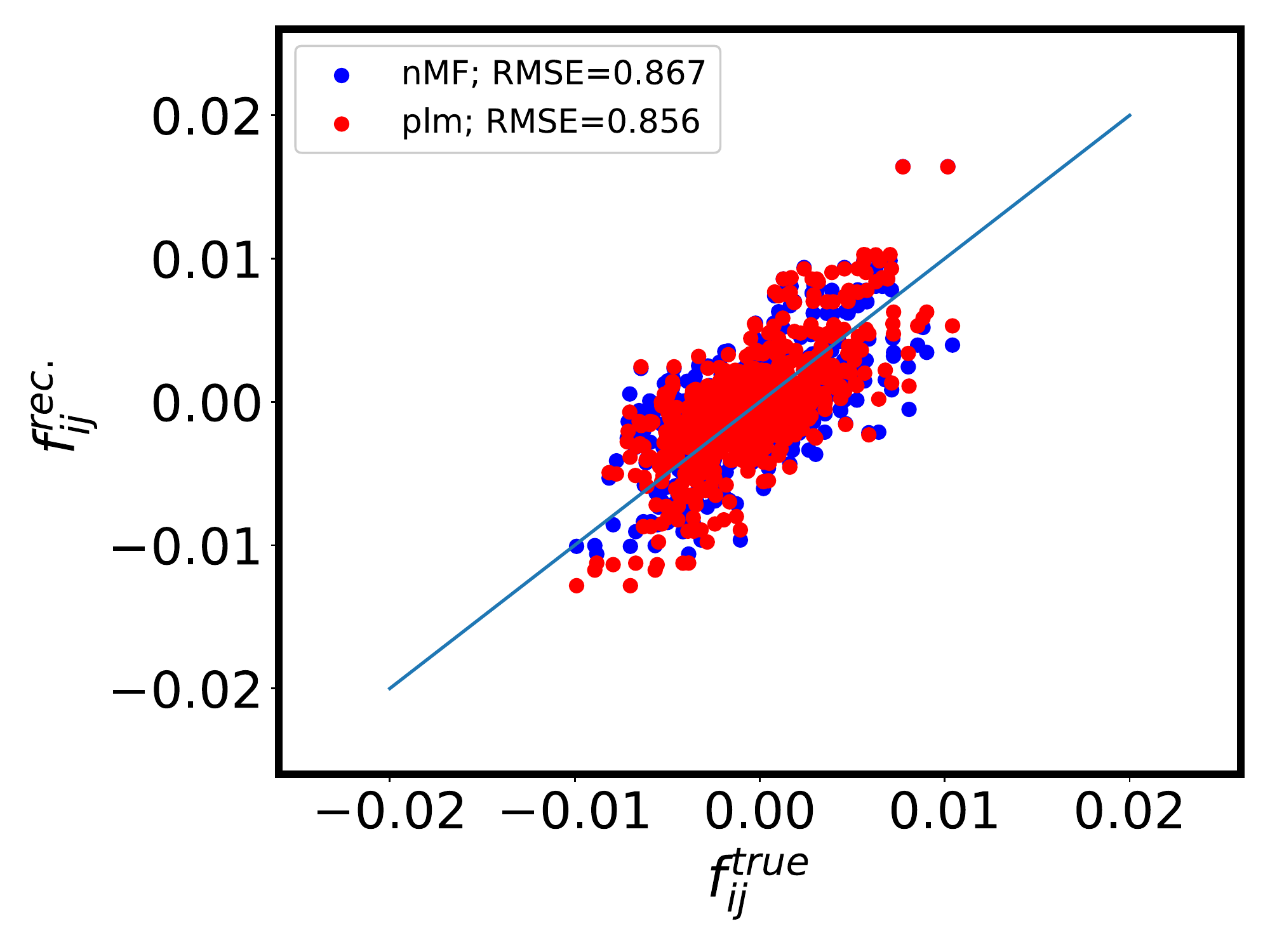}
\end{minipage}%
}
\caption{Examples of inference results for different recombination rates $r$
corresponding to three horizontally separated points in the middle of
Fig.~\protect\ref{fig:phase-diagram-mutation-vs-recombination}.
Mutation rate $\mu=0.05$, cross over rate $\rho=0.5$, fitness variation $\sigma=0.002$.
Ising parameters $J^*_{ij}$ inferred by the \textit{alltime-nMF} and \textit{alltime-PLM} procedures,
fitness values inferred from \protect\eqref{KNS-eq2}.
Panel (a) $r = 0.05$  shows that \protect\eqref{KNS-eq2} underestimates fitness, through there appears to be a functional relation.
Panel (b)  $r=0.5$ shows that both procedures estimate the fitness fairly accurately.
Panel (c)  $r=1.0$ shows KNS works worse compared with that for mediate $r$ presented in Panel (b).
}\label{fig:different-rec.-rate}
\end{figure}

\begin{figure}[!ht]
\centering
\subfigure[nMF]{
\begin{minipage}[t]{0.48\linewidth}
\centering
\includegraphics[width=\textwidth]{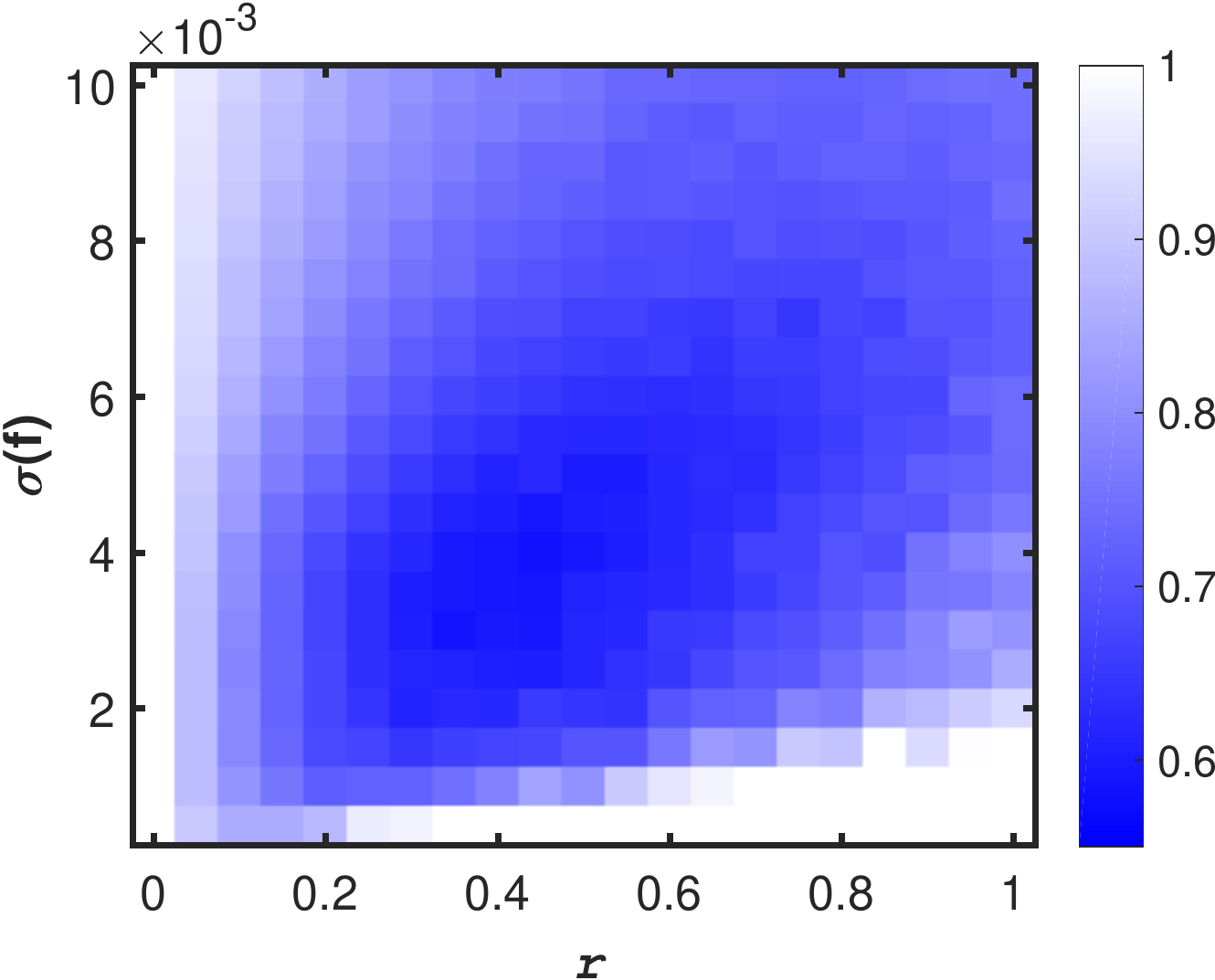}
\end{minipage}%
}
\subfigure[PLM]{
\begin{minipage}[t]{0.48\linewidth}
\centering
\includegraphics[width=\textwidth]{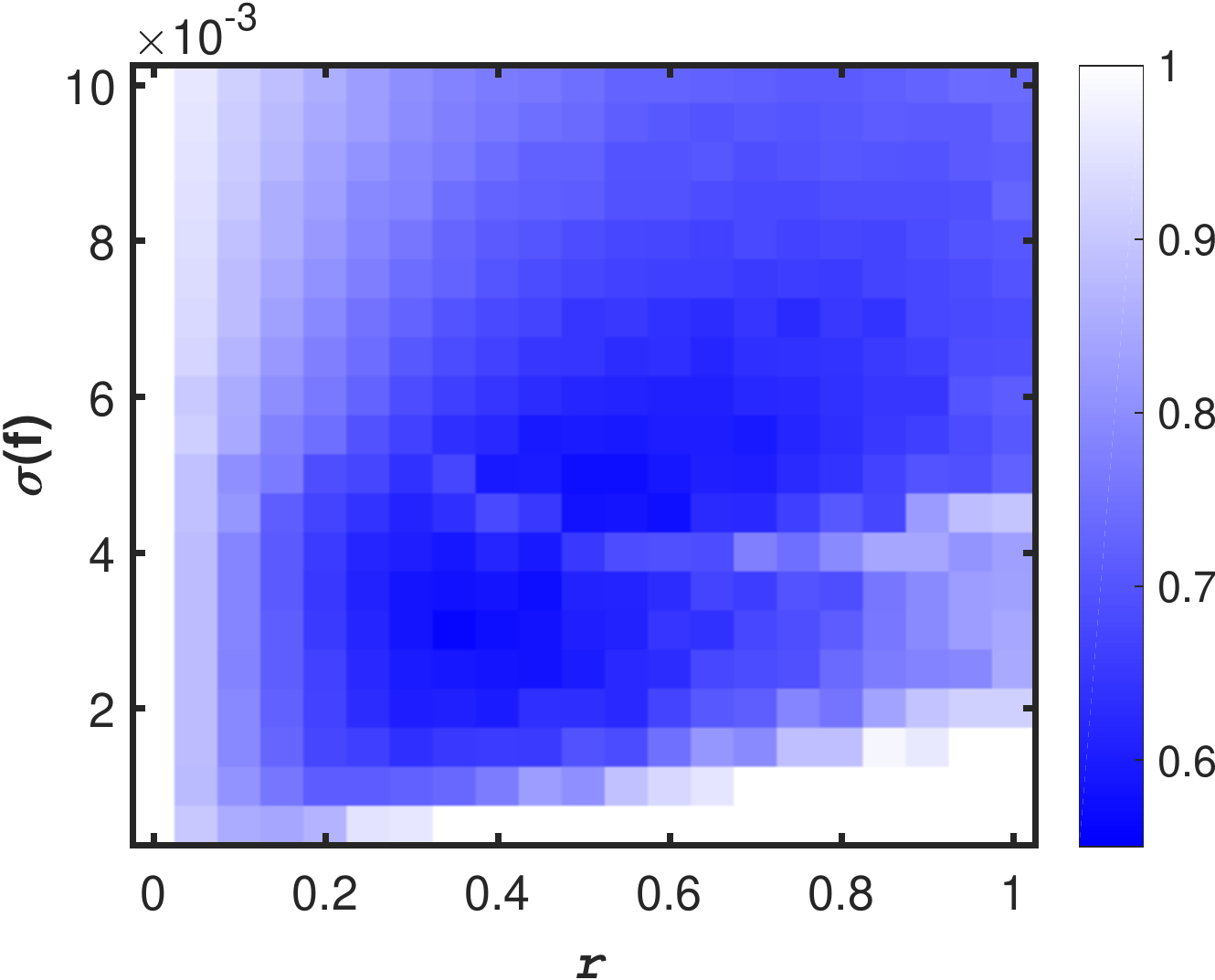}
\end{minipage}%
}
\caption{Phase diagram of epistatic fitness inference with parameters fitness $\sigma$ and recombination $r$.
The mutation rate $\mu=0.05$, cross over rate $\rho=0.5$.
The inference procedures are the same with that for Fig.~\ref{fig:phase-diagram-mutation-vs-recombination}.
By the color coding two schemes work in a very broad range of parameters.
For large recombination and very small variations
inference does not work.
}\label{fig:phase-diagram-fitness-vs-recombination}
\end{figure}

\begin{figure}[!ht]
\centering
\subfigure[$\sigma = 0.002$]{
\begin{minipage}[t]{0.3\linewidth}
\centering
\includegraphics[width=\textwidth]{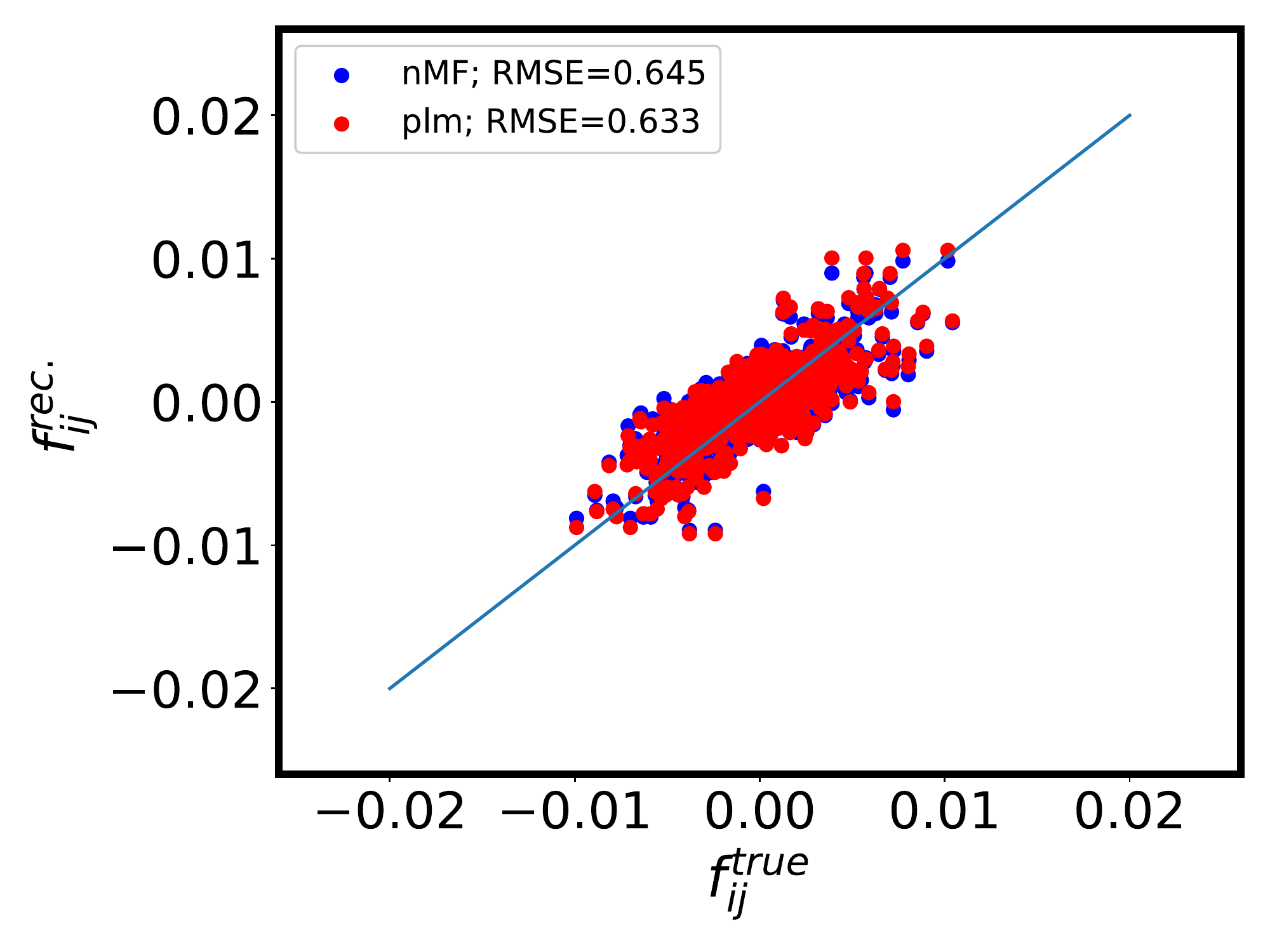}
\end{minipage}%
}
\subfigure[$\sigma = 0.008$]{
\begin{minipage}[t]{0.3\linewidth}
\centering
\includegraphics[width=\textwidth]{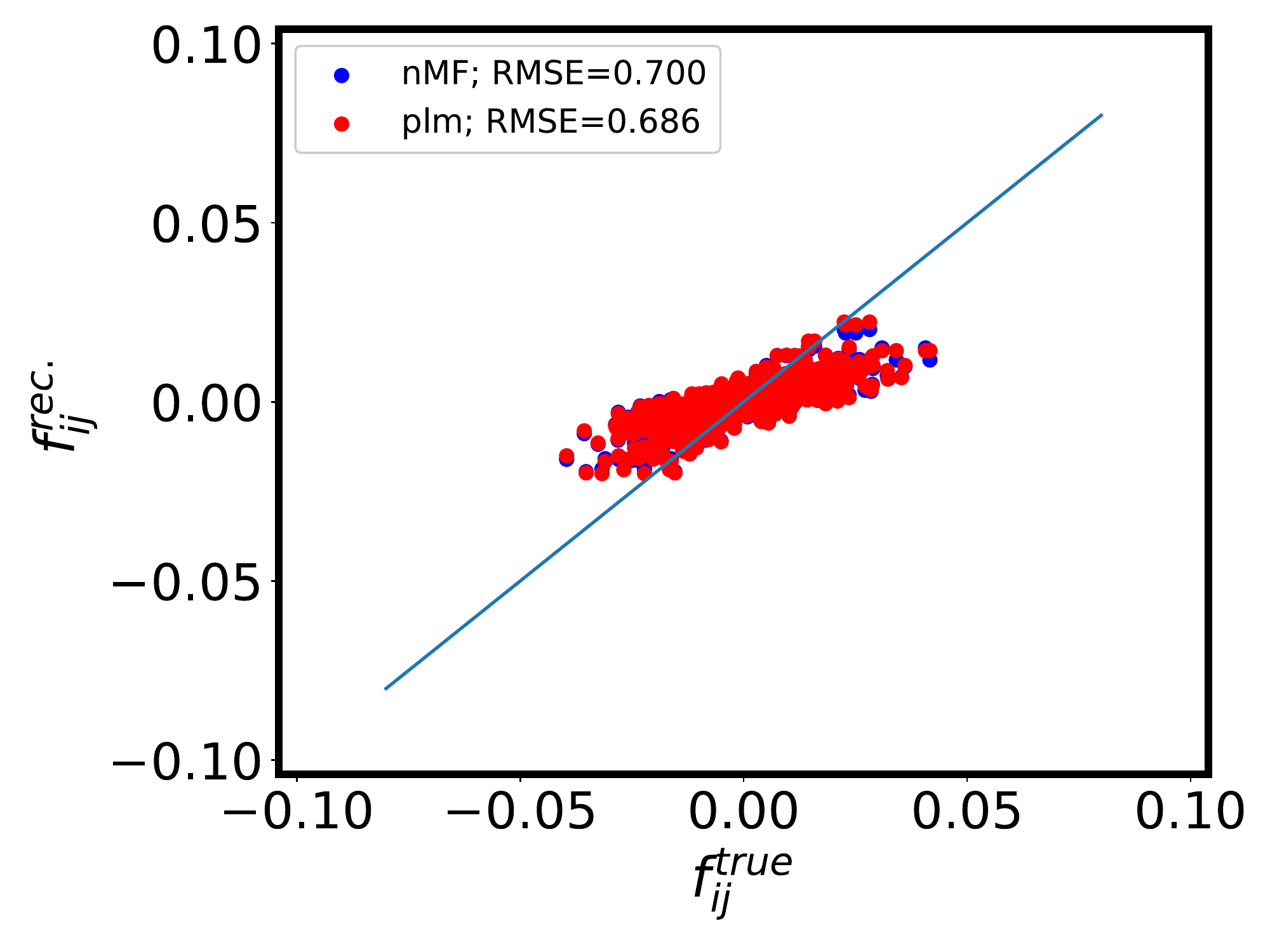}
\end{minipage}%
}
\subfigure[$\sigma = 0.02$]{
\begin{minipage}[t]{0.3\linewidth}
\centering
\includegraphics[width=\textwidth]{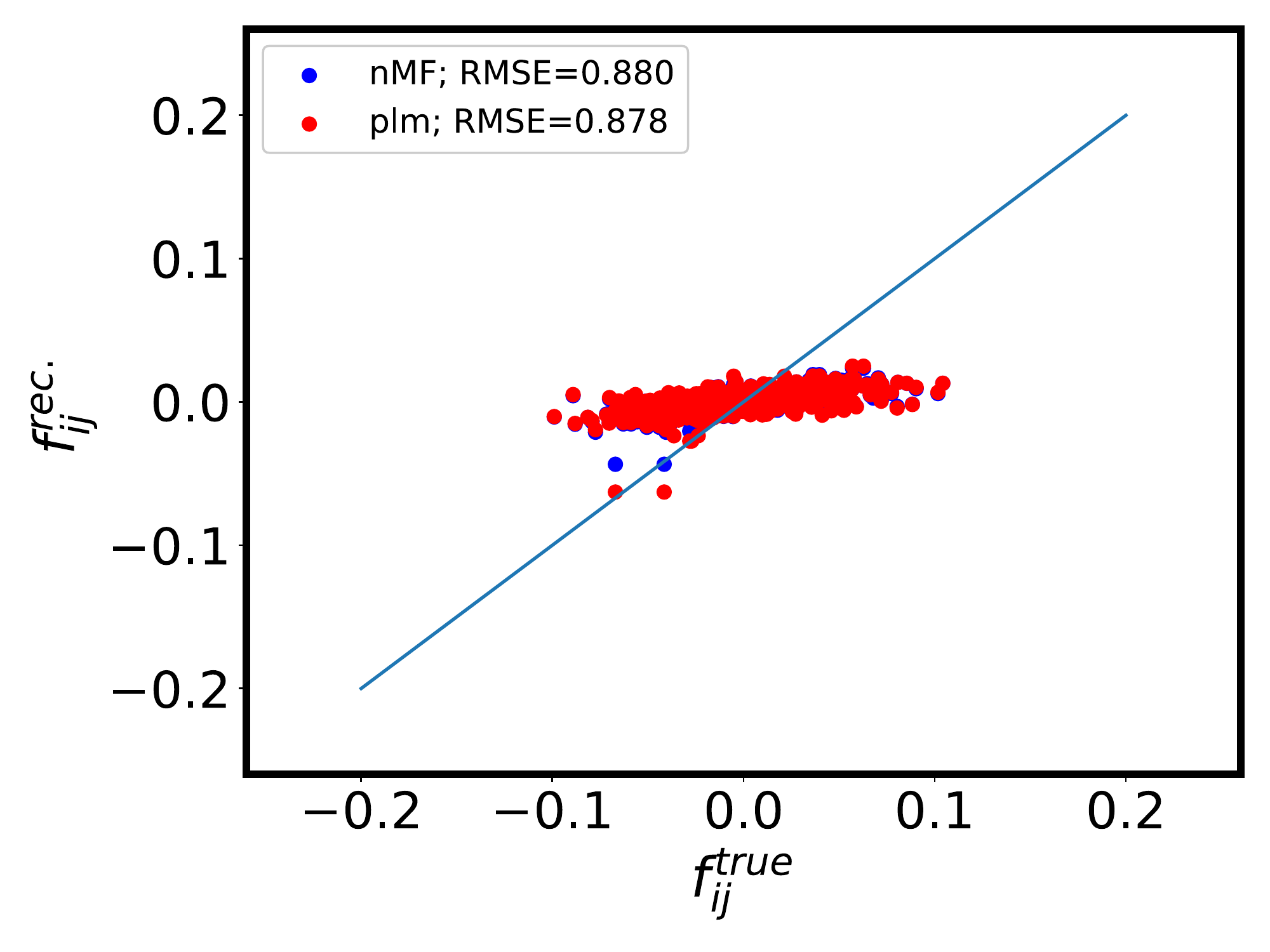}
\end{minipage}%
}
\caption{ Examples of inference results for different fitness variations $\sigma$ with recombination rate $r=0.5$
corresponding to three vertically separated points through the middle of
Fig.~\protect\ref{fig:phase-diagram-fitness-vs-recombination}.
Mutation rate $\mu= 0.05$, cross over rate $\rho = 0.5$.
Panel (a): $\sigma= 0.002$ shows  both procedures estimate the fitness correctly.
Panel (b): $\sigma= 0.008$ shows that \protect\eqref{KNS-eq2} underestimates fitness, through there appears to be a functional relation
with less scatter than in Panel (a).
Panel (c): $\sigma= 0.02$ (beyond the vertical range of Fig.~\protect\ref{fig:phase-diagram-fitness-vs-recombination})
shows that inference from \protect\eqref{KNS-eq2}
is approximately flat \textit{i.e.} uninformative.
}\label{fig:different-fitness-strength}
\end{figure}

Fig.~\ref{fig:phase-diagram-fitness-vs-recombination} similarly shows $\epsilon$
in the plane of fitness variations vs. recombination rate.
Also here inference works using formula \eqref{KNS-eq2} at intermediate values of mutation rate $\mu=0.05$.
Qualitatively this corresponds to the setting of Fig.~\ref{fig:scatter-plots}b
where $\epsilon $ is less than one.
For small recombination and large recombination we would not expect inference to work for the reasons as discussed above.
Additionally,  in the simulations reported in Fig.~\ref{fig:phase-diagram-fitness-vs-recombination},
apparently the fitness variability was not high enough for the last effect to show up.
This is confirmed by Fig.~\ref{fig:different-fitness-strength}, which shows the inference works worse with increasing fitness variations.

\section{Discussion}
\label{sec:discussion}
In this paper we have presented numerical evidence that
pair-wise fitness parameters in an evolving population
can be recovered from a distribution over genotypes.
Conceptually this allows to integrate the Direct Coupling Analysis (DCA) technique \cite{Stein-2015a,Michel-2017a,Cocco-2018a}
into the classic framework of population
genetics~\cite{Fisher-book,Blythe2007}.
We hope this will lead to further studies into when and
how such a connection can be made, and how to make it more precise.

An important conclusion of this work is that while recovery is possible in broad
parameter ranges it is not universally so. In this sense the basis for DCA
proposed here is quite different from max-entropy arguments.
For this reason we summarize the main regions where recovery is \textit{not} possible.

First, recovery is not possible for large recombination rate and small fitness variations.
This is so because in this range the $J_{ij}$ parameters which are inferred by DCA are small
and subject to small-sample noise.
This limitation is analogous to the limitations
on DCA when applied to finite data drawn from an Ising/Potts distribution, and which have
been investigated in the past by many groups~\cite{Nguyen-2017a}.

Second, for a given population size ($N$) and given simulation/observation time ($T$)
recovery is not possible if mutation rate ($\mu$) is low enough.
This is so because for such populations the structure is essentially
frozen. The underlying equations contain well-defined epistatic effects
but those are not reflected in the population structure over the time $T$, and hence cannot be recovered from population data.
Although of different origin this is similar to the limitations
on DCA when applied to data drawn from the Sherrington-Kirkpatrick model in the
low-temperature (spin-glass) phase, well investigated in the literature~\cite{Nguyen-2017a}.
In that latter case thermalization is slow and
many Monte Carlo updates are needed before two samples become uncorrelated;
the simulation/observation time ($T$) plays a similar role in our setting.

Third,
recovery is not possible when mutation rate is comparable to or larger than recombination rate.
In the data presented in this paper this
is exemplified by Figs.~\ref{fig:scatter-plots}c,~\ref{fig:different-rec.-rate}a,~\ref{fig:different-fitness-strength}b
and~\ref{fig:different-fitness-strength}c.
This discrepancy has no analogue in inverse Ising studies, and is instead,
we believe, because in these parameter ranges the Quasi-Linkage Equilibrium
theory of Kimura, Neher and Shraiman~\cite{Kimura1956,Kimura1964,Kimura1965,NeherShraiman2009, NeherShraiman2011}
is not applicable.
In contract to when recombination dominates over mutations there is no argument known to
us for what the population structure should then be.
It could be something very different from a Ising/Potts distribution,
but it could also be an Ising/Potts distribution with other relations among the parameters.
We note that especially Fig~\ref{fig:different-fitness-strength}b appears
to point a relation between fitness and the $J_{ij}$'s, but which is not well captured
QLE inference formula of \eqref{KNS-eq2}.
In any case, to understand the connection between epistatic fitness ($f_{ij}$) and population structure
($J_{ij}$), if any, is an important and unsolved problem.
  We hope to be able to return to this problem
  in a future contribution.

Finally, we note that the whole analysis in this paper is based on the concept of fitness landscapes.
Fitness is thus here an inherent (and heritable) property of each individual reflected in its expected
number of offspring, and one can imagine such fitness to be optimized.
While in this class of models absolute fitness depends on the whole population structure,
the relative order of fitness of two individuals does not. Darwinian fitness and natural selection is
in contrast a wider concept
including also competition and cooperation.
The relation between fitness and population structure when such aspects of strategy and game theory
are dominant are clearly very interesting, but lie beyond the range of approaches considered here.

\section*{Acknowledgments}
We thank Simona Cocco, Eugenio Mauro, R\'emi Monasson and Guilhem Semerjian
for discussions and Richard Neher for the FFPopSim software package
and comments.
The work of HLZ was supported
by National Natural Science Foundation of China (11705097),
Natural Science Foundation of Jiangsu Province (BK20170895),
Natural Science Foundation for Colleges and Universities in Jiangsu Province (17KJB140015)
Jiangsu Government Scholarship for Overseas Studies of 2018,
and
Scientific Research Foundation of Nanjing University of Posts and Telecommunications (NY217013).
The work of EA was
partially supported by Foundation for Polish Science
through TEAM-NET project (contract no. POIR.04.04.00-00-17C1/18-00).

\appendix

\section{FFPopSim settings}
\label{a:FFPopSim}
The FFPopSim package was written by
Fabio Zanini and Richard Neher and simulates
an evolving population with biallelic loci
and additive as well as  pairwise epistatic fitness functions~\cite{FFPopSim}.
We here describe first default parameters
which have the same
values in all simulations reported in this paper,
and then how we have varied simulation parameters
to obtain the different figures.

\subsection{FFPopSim default parameters}\label{a:FFPopSim-defaults}
The default parameters of FFPopSim we used in the simulations are listed in Tab.~\ref{tab:1}. The value of them is the same for the results presented here.
\begin{table}[!h]
\centering
\caption{Main default parameters of FFPopSim used in the simulation.}
\begin{tabular}{ll}
        number of loci (L)              &  25              \\
        circular                        &  True           \\
        carrying capacity (N)           &   200             \\
        generation                      &   $500\times 5$               \\
        recombination model             &   CROSSOVERS      \\
        crossover rate   ($\rho$)       &    $0.5$           \\
        fitness additive(coefficients)  &  $0.0 $\\
         number of traits                &   1               \\
\end{tabular}
\label{tab:1}
\end{table}

\subsection{FFPopSim parameter varied}
\label{a:FFPopSim-varied}
The following  parameters in FFPopSim are varied in the presented inference for epistatic fitness
\begin{table}[!h]
\centering
\caption{Varied parameters of FFPopSim used in the simulation.}
\begin{tabular}{ll}
        outcrossing rate ($r$)          &   $[0. , ~1.0]$             \\
        mutation rate    ($\mu$)        &   $[0.005, ~0.1]$          \\
        fitness coefficients   & Gaussian random\\ & number with std. $\sigma \in$ $[0. 0005, ~0.01]$  \\
        initial genotypes  & binary random numbers\\
\end{tabular}
\label{tab:2}
\end{table}

\section{Naive mean-field use (nMF)}
\label{a:nMF}
The multi-locus evolution is done by the ``evolve()" function for each generation with FFPopSim.
The correlations $\chi_{ij}$s are computed and recorded for each generation.
When the ``evolve" process is done, the mean correlations $\left<\chi_{ij}\right>$ over generations are computed and used for the inference.
The pseudo-code for nMF inference is presented in Algorithm \ref{alg:nMF}.

\begin{algorithm}\label{alg:nMF}
\LinesNumbered
\caption{Epistatic fitness inference by nMF procedure: $f_{ij}^{nMF}$}
 \SetAlgoLined
 \setstretch{1.35}
   \KwIn{mean correlations: $\left<\chi_{ij}\right>$}
   \KwOut{inferred epistatic fitness: $f_{ij}^{nMF}$ }
\begin{algorithmic}[1]
       \STATE import \textbf{scipy}
       \STATE from \textbf{scipy} import \textbf{linalg}
       \STATE $J_{ij}^{nMF}$ = - linalg.inv($\left<\chi_{ij}\right>$)
       \STATE $f_{ij}^{nMF} = J_{ij}^{nMF} \ast r \ast c_{ij}$
\end{algorithmic}
\end{algorithm}

\section{Pseudo-likelihood maximization use (PLM)}
\label{a:PLM}
Ising model parameters are inferred by the Pseudo-likelihood maximization method \cite{Ekeberg-2014a} using the matlab software PLM at \cite{Gao-github} (www. github.com/gaochenyi/CC-PLM).
The allele states are recorded for each generation. A giant matrix with an approximate size of $25\times N\times T$ produced by FFPopSim is the input of PLM software for the Ising parameters $J_{ij}^{PLM}$.
The pseudo-code for PLM inference is Algorithm \ref{alg:PLM}.

With $L=25$ loci and the rest of parameters are the same, the CPU-time for $N=200$ and $T = 500$
on a standard desktop computer is about 50 seconds and it is dominated by the PLM process. It increases linearly with increasing $N$ or $T$. For instance, it will be around 3000 seconds for $N=2000$ and $T = 2000$, in which the PLM process costs about 2300 seconds.

\begin{algorithm}
\LinesNumbered
\caption{Epistatic fitness inference by PLM procedure: $f_{ij}^{PLM}$}
\label{alg:PLM}
 \SetAlgoLined
 \setstretch{1.35}
   \KwIn{Giant binary state matrix $\mathbf{S}$ with an approximate size of $L\times N\times T$}
   \KwOut{inferred epistatic fitness:  $f_{ij}^{PLM}$ }
\begin{algorithmic}[1]
       \STATE import \textbf{matlab.engine}
       \STATE eng = matlab.engine.start\_matlab()
       \STATE $J_{ij}^{PLM}$ = eng.plm\_to\_ffpopsim($\textbf{S}$)
       \STATE $f_{ij}^{PLM} = J_{ij}^{PLM} \ast r \ast c_{ij}$
\end{algorithmic}
\end{algorithm}

\bibliography{ref,ref2}

\begin{thebibliography}{32}%
\makeatletter
\providecommand \@ifxundefined [1]{%
 \@ifx{#1\undefined}
}%
\providecommand \@ifnum [1]{%
 \ifnum #1\expandafter \@firstoftwo
 \else \expandafter \@secondoftwo
 \fi
}%
\providecommand \@ifx [1]{%
 \ifx #1\expandafter \@firstoftwo
 \else \expandafter \@secondoftwo
 \fi
}%
\providecommand \natexlab [1]{#1}%
\providecommand \enquote  [1]{``#1''}%
\providecommand \bibnamefont  [1]{#1}%
\providecommand \bibfnamefont [1]{#1}%
\providecommand \citenamefont [1]{#1}%
\providecommand \href@noop [0]{\@secondoftwo}%
\providecommand \href [0]{\begingroup \@sanitize@url \@href}%
\providecommand \@href[1]{\@@startlink{#1}\@@href}%
\providecommand \@@href[1]{\endgroup#1\@@endlink}%
\providecommand \@sanitize@url [0]{\catcode `\\12\catcode `\$12\catcode
  `\&12\catcode `\#12\catcode `\^12\catcode `\_12\catcode `\%12\relax}%
\providecommand \@@startlink[1]{}%
\providecommand \@@endlink[0]{}%
\providecommand \url  [0]{\begingroup\@sanitize@url \@url }%
\providecommand \@url [1]{\endgroup\@href {#1}{\urlprefix }}%
\providecommand \urlprefix  [0]{URL }%
\providecommand \Eprint [0]{\href }%
\providecommand \doibase [0]{http://dx.doi.org/}%
\providecommand \selectlanguage [0]{\@gobble}%
\providecommand \bibinfo  [0]{\@secondoftwo}%
\providecommand \bibfield  [0]{\@secondoftwo}%
\providecommand \translation [1]{[#1]}%
\providecommand \BibitemOpen [0]{}%
\providecommand \bibitemStop [0]{}%
\providecommand \bibitemNoStop [0]{.\EOS\space}%
\providecommand \EOS [0]{\spacefactor3000\relax}%
\providecommand \BibitemShut  [1]{\csname bibitem#1\endcsname}%
\let\auto@bib@innerbib\@empty
\bibitem [{\citenamefont {Nguyen}\ \emph {et~al.}(2017)\citenamefont {Nguyen},
  \citenamefont {Zecchina},\ and\ \citenamefont {Berg}}]{Nguyen-2017a}%
  \BibitemOpen
  \bibfield  {author} {\bibinfo {author} {\bibfnamefont {H.~C.}\ \bibnamefont
  {Nguyen}}, \bibinfo {author} {\bibfnamefont {R.}~\bibnamefont {Zecchina}}, \
  and\ \bibinfo {author} {\bibfnamefont {J.}~\bibnamefont {Berg}},\ }\href
  {\doibase 10.1080/00018732.2017.1341604} {\bibfield  {journal} {\bibinfo
  {journal} {Adv. Phys.}\ }\textbf {\bibinfo {volume} {66}},\ \bibinfo {pages}
  {197} (\bibinfo {year} {2017})}\BibitemShut {NoStop}%
\bibitem [{\citenamefont {Weigt}\ \emph {et~al.}(2009)\citenamefont {Weigt},
  \citenamefont {White}, \citenamefont {Szurmant}, \citenamefont {Hoch},\ and\
  \citenamefont {Hwa}}]{Weigt-2009a}%
  \BibitemOpen
  \bibfield  {author} {\bibinfo {author} {\bibfnamefont {M.}~\bibnamefont
  {Weigt}}, \bibinfo {author} {\bibfnamefont {R.~A.}\ \bibnamefont {White}},
  \bibinfo {author} {\bibfnamefont {H.}~\bibnamefont {Szurmant}}, \bibinfo
  {author} {\bibfnamefont {J.~A.}\ \bibnamefont {Hoch}}, \ and\ \bibinfo
  {author} {\bibfnamefont {T.}~\bibnamefont {Hwa}},\ }\href {\doibase
  10.1073/pnas.0805923106} {\bibfield  {journal} {\bibinfo  {journal} {Proc.
  Natl. Acad. Sci.}\ }\textbf {\bibinfo {volume} {106}},\ \bibinfo {pages} {67}
  (\bibinfo {year} {2009})}\BibitemShut {NoStop}%
\bibitem [{\citenamefont {Morcos}\ \emph {et~al.}(2011)\citenamefont {Morcos},
  \citenamefont {Pagnani}, \citenamefont {Lunt}, \citenamefont {Bertolino},
  \citenamefont {Marks}, \citenamefont {Sander}, \citenamefont {Zecchina},
  \citenamefont {Onuchic}, \citenamefont {Hwa},\ and\ \citenamefont
  {Weigt}}]{Morcos-2011a}%
  \BibitemOpen
  \bibfield  {author} {\bibinfo {author} {\bibfnamefont {F.}~\bibnamefont
  {Morcos}}, \bibinfo {author} {\bibfnamefont {A.}~\bibnamefont {Pagnani}},
  \bibinfo {author} {\bibfnamefont {B.}~\bibnamefont {Lunt}}, \bibinfo {author}
  {\bibfnamefont {A.}~\bibnamefont {Bertolino}}, \bibinfo {author}
  {\bibfnamefont {D.~S.}\ \bibnamefont {Marks}}, \bibinfo {author}
  {\bibfnamefont {C.}~\bibnamefont {Sander}}, \bibinfo {author} {\bibfnamefont
  {R.}~\bibnamefont {Zecchina}}, \bibinfo {author} {\bibfnamefont {J.~N.}\
  \bibnamefont {Onuchic}}, \bibinfo {author} {\bibfnamefont {T.}~\bibnamefont
  {Hwa}}, \ and\ \bibinfo {author} {\bibfnamefont {M.}~\bibnamefont {Weigt}},\
  }\href {\doibase 10.1073/pnas.1111471108} {\bibfield  {journal} {\bibinfo
  {journal} {Proc. Natl. Acad. Sci.}\ }\textbf {\bibinfo {volume} {108}},\
  \bibinfo {pages} {E1293} (\bibinfo {year} {2011})}\BibitemShut {NoStop}%
\bibitem [{\citenamefont {Stein}\ \emph {et~al.}(2015)\citenamefont {Stein},
  \citenamefont {Marks},\ and\ \citenamefont {Sander}}]{Stein-2015a}%
  \BibitemOpen
  \bibfield  {author} {\bibinfo {author} {\bibfnamefont {R.~R.}\ \bibnamefont
  {Stein}}, \bibinfo {author} {\bibfnamefont {D.~S.}\ \bibnamefont {Marks}}, \
  and\ \bibinfo {author} {\bibfnamefont {C.}~\bibnamefont {Sander}},\ }\href
  {\doibase 10.1371/journal.pcbi.1004182} {\bibfield  {journal} {\bibinfo
  {journal} {PLoS Comput. Biol.}\ }\textbf {\bibinfo {volume} {11}},\ \bibinfo
  {pages} {e1004182} (\bibinfo {year} {2015})}\BibitemShut {NoStop}%
\bibitem [{\citenamefont {Michel}\ \emph
  {et~al.}(2017{\natexlab{a}})\citenamefont {Michel}, \citenamefont {Skwark},
  \citenamefont {Men{\'e}ndez~Hurtado}, \citenamefont {Ekeberg},\ and\
  \citenamefont {Elofsson}}]{Michel-2017a}%
  \BibitemOpen
  \bibfield  {author} {\bibinfo {author} {\bibfnamefont {M.}~\bibnamefont
  {Michel}}, \bibinfo {author} {\bibfnamefont {M.~J.}\ \bibnamefont {Skwark}},
  \bibinfo {author} {\bibfnamefont {D.}~\bibnamefont {Men{\'e}ndez~Hurtado}},
  \bibinfo {author} {\bibfnamefont {M.}~\bibnamefont {Ekeberg}}, \ and\
  \bibinfo {author} {\bibfnamefont {A.}~\bibnamefont {Elofsson}},\ }\href
  {\doibase 10.1093/bioinformatics/btx332} {\bibfield  {journal} {\bibinfo
  {journal} {Bioinformatics}\ }\textbf {\bibinfo {volume} {33}},\ \bibinfo
  {pages} {2859} (\bibinfo {year} {2017}{\natexlab{a}})}\BibitemShut {NoStop}%
\bibitem [{\citenamefont {Cocco}\ \emph {et~al.}(2018)\citenamefont {Cocco},
  \citenamefont {Feinauer}, \citenamefont {Figliuzzi}, \citenamefont
  {Monasson},\ and\ \citenamefont {Weigt}}]{Cocco-2018a}%
  \BibitemOpen
  \bibfield  {author} {\bibinfo {author} {\bibfnamefont {S.}~\bibnamefont
  {Cocco}}, \bibinfo {author} {\bibfnamefont {C.}~\bibnamefont {Feinauer}},
  \bibinfo {author} {\bibfnamefont {M.}~\bibnamefont {Figliuzzi}}, \bibinfo
  {author} {\bibfnamefont {R.}~\bibnamefont {Monasson}}, \ and\ \bibinfo
  {author} {\bibfnamefont {M.}~\bibnamefont {Weigt}},\ }\href {\doibase
  10.1088/1361-6633/aa9965} {\bibfield  {journal} {\bibinfo  {journal} {Rep.
  Prog. Phys.}\ }\textbf {\bibinfo {volume} {81}},\ \bibinfo {pages} {032601}
  (\bibinfo {year} {2018})}\BibitemShut {NoStop}%
\bibitem [{\citenamefont {Ovchinnikov}\ \emph {et~al.}(2017)\citenamefont
  {Ovchinnikov}, \citenamefont {Park}, \citenamefont {Varghese}, \citenamefont
  {Huang}, \citenamefont {Pavlopoulos}, \citenamefont {Kim}, \citenamefont
  {Kamisetty}, \citenamefont {Kyrpides},\ and\ \citenamefont
  {Baker}}]{Ovchinnikov-2017a}%
  \BibitemOpen
  \bibfield  {author} {\bibinfo {author} {\bibfnamefont {S.}~\bibnamefont
  {Ovchinnikov}}, \bibinfo {author} {\bibfnamefont {H.}~\bibnamefont {Park}},
  \bibinfo {author} {\bibfnamefont {N.}~\bibnamefont {Varghese}}, \bibinfo
  {author} {\bibfnamefont {P.-S.}\ \bibnamefont {Huang}}, \bibinfo {author}
  {\bibfnamefont {G.~A.}\ \bibnamefont {Pavlopoulos}}, \bibinfo {author}
  {\bibfnamefont {D.~E.}\ \bibnamefont {Kim}}, \bibinfo {author} {\bibfnamefont
  {H.}~\bibnamefont {Kamisetty}}, \bibinfo {author} {\bibfnamefont {N.~C.}\
  \bibnamefont {Kyrpides}}, \ and\ \bibinfo {author} {\bibfnamefont
  {D.}~\bibnamefont {Baker}},\ }\href {\doibase 10.1126/science.aah4043}
  {\bibfield  {journal} {\bibinfo  {journal} {Science}\ }\textbf {\bibinfo
  {volume} {355}},\ \bibinfo {pages} {294} (\bibinfo {year}
  {2017})}\BibitemShut {NoStop}%
\bibitem [{\citenamefont {Michel}\ \emph
  {et~al.}(2017{\natexlab{b}})\citenamefont {Michel}, \citenamefont
  {Men{\'e}ndez~Hurtado}, \citenamefont {Uziela},\ and\ \citenamefont
  {Elofsson}}]{Michel-2017b}%
  \BibitemOpen
  \bibfield  {author} {\bibinfo {author} {\bibfnamefont {M.}~\bibnamefont
  {Michel}}, \bibinfo {author} {\bibfnamefont {D.}~\bibnamefont
  {Men{\'e}ndez~Hurtado}}, \bibinfo {author} {\bibfnamefont {K.}~\bibnamefont
  {Uziela}}, \ and\ \bibinfo {author} {\bibfnamefont {A.}~\bibnamefont
  {Elofsson}},\ }\href {\doibase 10.1093/bioinformatics/btx239} {\bibfield
  {journal} {\bibinfo  {journal} {Bioinformatics}\ }\textbf {\bibinfo {volume}
  {33}},\ \bibinfo {pages} {i23} (\bibinfo {year}
  {2017}{\natexlab{b}})}\BibitemShut {NoStop}%
\bibitem [{\citenamefont {Ovchinnikov}\ \emph {et~al.}(2018)\citenamefont
  {Ovchinnikov}, \citenamefont {Park}, \citenamefont {Kim}, \citenamefont
  {DiMaio},\ and\ \citenamefont {Baker}}]{Ovchinnikov-2018a}%
  \BibitemOpen
  \bibfield  {author} {\bibinfo {author} {\bibfnamefont {S.}~\bibnamefont
  {Ovchinnikov}}, \bibinfo {author} {\bibfnamefont {H.}~\bibnamefont {Park}},
  \bibinfo {author} {\bibfnamefont {D.~E.}\ \bibnamefont {Kim}}, \bibinfo
  {author} {\bibfnamefont {F.}~\bibnamefont {DiMaio}}, \ and\ \bibinfo {author}
  {\bibfnamefont {D.}~\bibnamefont {Baker}},\ }\href {\doibase
  10.1002/prot.25390} {\bibfield  {journal} {\bibinfo  {journal} {Proteins}\
  }\textbf {\bibinfo {volume} {86}},\ \bibinfo {pages} {113} (\bibinfo {year}
  {2018})}\BibitemShut {NoStop}%
\bibitem [{\citenamefont {De~Leonardis}\ \emph {et~al.}(2015)\citenamefont
  {De~Leonardis}, \citenamefont {Lutz}, \citenamefont {Ratz}, \citenamefont
  {Simona}, \citenamefont {Monasson}, \citenamefont {Weigt},\ and\
  \citenamefont {Schug}}]{DeLeonardis-2015a}%
  \BibitemOpen
  \bibfield  {author} {\bibinfo {author} {\bibfnamefont {E.}~\bibnamefont
  {De~Leonardis}}, \bibinfo {author} {\bibfnamefont {B.}~\bibnamefont {Lutz}},
  \bibinfo {author} {\bibfnamefont {S.}~\bibnamefont {Ratz}}, \bibinfo {author}
  {\bibfnamefont {C.}~\bibnamefont {Simona}}, \bibinfo {author} {\bibfnamefont
  {R.}~\bibnamefont {Monasson}}, \bibinfo {author} {\bibfnamefont
  {M.}~\bibnamefont {Weigt}}, \ and\ \bibinfo {author} {\bibfnamefont
  {A.}~\bibnamefont {Schug}},\ }\href {\doibase 10.1093/nar/gkv932} {\bibfield
  {journal} {\bibinfo  {journal} {Nucleic Acids Res.}\ }\textbf {\bibinfo
  {volume} {43}},\ \bibinfo {pages} {10444} (\bibinfo {year}
  {2015})}\BibitemShut {NoStop}%
\bibitem [{\citenamefont {Gueudr\'e}\ \emph {et~al.}(2016)\citenamefont
  {Gueudr\'e}, \citenamefont {Baldassi}, \citenamefont {Zamparo}, \citenamefont
  {Weigt},\ and\ \citenamefont {Pagnani}}]{Gueudre-2016a}%
  \BibitemOpen
  \bibfield  {author} {\bibinfo {author} {\bibfnamefont {T.}~\bibnamefont
  {Gueudr\'e}}, \bibinfo {author} {\bibfnamefont {C.}~\bibnamefont {Baldassi}},
  \bibinfo {author} {\bibfnamefont {M.}~\bibnamefont {Zamparo}}, \bibinfo
  {author} {\bibfnamefont {M.}~\bibnamefont {Weigt}}, \ and\ \bibinfo {author}
  {\bibfnamefont {A.}~\bibnamefont {Pagnani}},\ }\href {\doibase
  10.1073/pnas.1607570113} {\bibfield  {journal} {\bibinfo  {journal} {Proc.
  Natl. Acad. Sci.}\ }\textbf {\bibinfo {volume} {113}},\ \bibinfo {pages}
  {12186} (\bibinfo {year} {2016})}\BibitemShut {NoStop}%
\bibitem [{\citenamefont {Uguzzoni}\ \emph {et~al.}(2017)\citenamefont
  {Uguzzoni}, \citenamefont {John~Lovis}, \citenamefont {Oteri}, \citenamefont
  {Schug}, \citenamefont {Szurmant},\ and\ \citenamefont
  {Weigt}}]{Uguzzoni-2017a}%
  \BibitemOpen
  \bibfield  {author} {\bibinfo {author} {\bibfnamefont {G.}~\bibnamefont
  {Uguzzoni}}, \bibinfo {author} {\bibfnamefont {S.}~\bibnamefont
  {John~Lovis}}, \bibinfo {author} {\bibfnamefont {F.}~\bibnamefont {Oteri}},
  \bibinfo {author} {\bibfnamefont {A.}~\bibnamefont {Schug}}, \bibinfo
  {author} {\bibfnamefont {H.}~\bibnamefont {Szurmant}}, \ and\ \bibinfo
  {author} {\bibfnamefont {M.}~\bibnamefont {Weigt}},\ }\href {\doibase
  10.1073/pnas.1615068114} {\bibfield  {journal} {\bibinfo  {journal} {Proc.
  Natl. Acad. Sci.}\ }\textbf {\bibinfo {volume} {114}},\ \bibinfo {pages}
  {E2662} (\bibinfo {year} {2017})}\BibitemShut {NoStop}%
\bibitem [{\citenamefont {Weinreb}\ \emph {et~al.}(2016)\citenamefont
  {Weinreb}, \citenamefont {Riesselman}, \citenamefont {Ingraham},
  \citenamefont {Gross}, \citenamefont {Sander},\ and\ \citenamefont
  {Marks}}]{Weinreb-2016a}%
  \BibitemOpen
  \bibfield  {author} {\bibinfo {author} {\bibfnamefont {C.}~\bibnamefont
  {Weinreb}}, \bibinfo {author} {\bibfnamefont {A.~J.}\ \bibnamefont
  {Riesselman}}, \bibinfo {author} {\bibfnamefont {J.~B.}\ \bibnamefont
  {Ingraham}}, \bibinfo {author} {\bibfnamefont {T.}~\bibnamefont {Gross}},
  \bibinfo {author} {\bibfnamefont {C.}~\bibnamefont {Sander}}, \ and\ \bibinfo
  {author} {\bibfnamefont {D.~S.}\ \bibnamefont {Marks}},\ }\href {\doibase
  10.1016/j.cell.2016.03.030} {\bibfield  {journal} {\bibinfo  {journal}
  {Cell}\ }\textbf {\bibinfo {volume} {165}},\ \bibinfo {pages} {963} (\bibinfo
  {year} {2016})}\BibitemShut {NoStop}%
\bibitem [{\citenamefont {Figliuzzi}\ \emph {et~al.}(2016)\citenamefont
  {Figliuzzi}, \citenamefont {Jacquier}, \citenamefont {Schug}, \citenamefont
  {Tenaillon},\ and\ \citenamefont {Weigt}}]{Figliuzzi-2016a}%
  \BibitemOpen
  \bibfield  {author} {\bibinfo {author} {\bibfnamefont {M.}~\bibnamefont
  {Figliuzzi}}, \bibinfo {author} {\bibfnamefont {H.}~\bibnamefont {Jacquier}},
  \bibinfo {author} {\bibfnamefont {A.}~\bibnamefont {Schug}}, \bibinfo
  {author} {\bibfnamefont {O.}~\bibnamefont {Tenaillon}}, \ and\ \bibinfo
  {author} {\bibfnamefont {M.}~\bibnamefont {Weigt}},\ }\href {\doibase
  10.1093/molbev/msv211} {\bibfield  {journal} {\bibinfo  {journal} {Mol. Biol.
  Evol.}\ }\textbf {\bibinfo {volume} {33}},\ \bibinfo {pages} {268} (\bibinfo
  {year} {2016})}\BibitemShut {NoStop}%
\bibitem [{\citenamefont {Hopf}\ \emph {et~al.}(2017)\citenamefont {Hopf},
  \citenamefont {Ingraham}, \citenamefont {Poelwijk}, \citenamefont {Scharfe},
  \citenamefont {Springer}, \citenamefont {Sander},\ and\ \citenamefont
  {Marks}}]{Hopf-2017a}%
  \BibitemOpen
  \bibfield  {author} {\bibinfo {author} {\bibfnamefont {T.~A.}\ \bibnamefont
  {Hopf}}, \bibinfo {author} {\bibfnamefont {J.~B.}\ \bibnamefont {Ingraham}},
  \bibinfo {author} {\bibfnamefont {F.~J.}\ \bibnamefont {Poelwijk}}, \bibinfo
  {author} {\bibfnamefont {C.~P.~I.}\ \bibnamefont {Scharfe}}, \bibinfo
  {author} {\bibfnamefont {M.}~\bibnamefont {Springer}}, \bibinfo {author}
  {\bibfnamefont {C.}~\bibnamefont {Sander}}, \ and\ \bibinfo {author}
  {\bibfnamefont {D.~S.}\ \bibnamefont {Marks}},\ }\href {\doibase
  10.1038/nbt.3769} {\bibfield  {journal} {\bibinfo  {journal} {Nat.
  Biotechnol.}\ }\textbf {\bibinfo {volume} {35}},\ \bibinfo {pages} {128}
  (\bibinfo {year} {2017})}\BibitemShut {NoStop}%
\bibitem [{\citenamefont {Couce}\ \emph {et~al.}(2017)\citenamefont {Couce},
  \citenamefont {Caudwell}, \citenamefont {Feinauer}, \citenamefont
  {Hindr{\'e}}, \citenamefont {Feugeas}, \citenamefont {Weigt}, \citenamefont
  {Lenski}, \citenamefont {Schneider},\ and\ \citenamefont
  {Tenaillon}}]{CouceE9026}%
  \BibitemOpen
  \bibfield  {author} {\bibinfo {author} {\bibfnamefont {A.}~\bibnamefont
  {Couce}}, \bibinfo {author} {\bibfnamefont {L.~V.}\ \bibnamefont {Caudwell}},
  \bibinfo {author} {\bibfnamefont {C.}~\bibnamefont {Feinauer}}, \bibinfo
  {author} {\bibfnamefont {T.}~\bibnamefont {Hindr{\'e}}}, \bibinfo {author}
  {\bibfnamefont {J.-P.}\ \bibnamefont {Feugeas}}, \bibinfo {author}
  {\bibfnamefont {M.}~\bibnamefont {Weigt}}, \bibinfo {author} {\bibfnamefont
  {R.~E.}\ \bibnamefont {Lenski}}, \bibinfo {author} {\bibfnamefont
  {D.}~\bibnamefont {Schneider}}, \ and\ \bibinfo {author} {\bibfnamefont
  {O.}~\bibnamefont {Tenaillon}},\ }\href {\doibase 10.1073/pnas.1705887114}
  {\bibfield  {journal} {\bibinfo  {journal} {Proc. Natl. Acad. Sci.}\ }\textbf
  {\bibinfo {volume} {114}},\ \bibinfo {pages} {E9026} (\bibinfo {year}
  {2017})}\BibitemShut {NoStop}%
\bibitem [{\citenamefont {Skwark}\ \emph {et~al.}(2017)\citenamefont {Skwark},
  \citenamefont {Croucher}, \citenamefont {Puranen}, \citenamefont
  {Chewapreecha}, \citenamefont {Pesonen}, \citenamefont {Xu}, \citenamefont
  {Turner}, \citenamefont {Harris}, \citenamefont {Beres}, \citenamefont
  {Musser}, \citenamefont {Parkhill}, \citenamefont {Bentley}, \citenamefont
  {Aurell},\ and\ \citenamefont {Corander}}]{Skwark-2017a}%
  \BibitemOpen
  \bibfield  {author} {\bibinfo {author} {\bibfnamefont {M.~J.}\ \bibnamefont
  {Skwark}}, \bibinfo {author} {\bibfnamefont {N.~J.}\ \bibnamefont
  {Croucher}}, \bibinfo {author} {\bibfnamefont {S.}~\bibnamefont {Puranen}},
  \bibinfo {author} {\bibfnamefont {C.}~\bibnamefont {Chewapreecha}}, \bibinfo
  {author} {\bibfnamefont {M.}~\bibnamefont {Pesonen}}, \bibinfo {author}
  {\bibfnamefont {Y.~Y.}\ \bibnamefont {Xu}}, \bibinfo {author} {\bibfnamefont
  {P.}~\bibnamefont {Turner}}, \bibinfo {author} {\bibfnamefont {S.~R.}\
  \bibnamefont {Harris}}, \bibinfo {author} {\bibfnamefont {S.~B.}\
  \bibnamefont {Beres}}, \bibinfo {author} {\bibfnamefont {J.~M.}\ \bibnamefont
  {Musser}}, \bibinfo {author} {\bibfnamefont {J.}~\bibnamefont {Parkhill}},
  \bibinfo {author} {\bibfnamefont {S.~D.}\ \bibnamefont {Bentley}}, \bibinfo
  {author} {\bibfnamefont {E.}~\bibnamefont {Aurell}}, \ and\ \bibinfo {author}
  {\bibfnamefont {J.}~\bibnamefont {Corander}},\ }\href {\doibase
  10.1371/journal.pgen.1006508} {\bibfield  {journal} {\bibinfo  {journal}
  {PLos Genet.}\ }\textbf {\bibinfo {volume} {13}},\ \bibinfo {pages}
  {e1006508} (\bibinfo {year} {2017})}\BibitemShut {NoStop}%
\bibitem [{\citenamefont {Schubert}\ \emph {et~al.}(2019)\citenamefont
  {Schubert}, \citenamefont {Maddamsetti}, \citenamefont {Nyman}, \citenamefont
  {Farhat},\ and\ \citenamefont {Marks}}]{Schubert2019}%
  \BibitemOpen
  \bibfield  {author} {\bibinfo {author} {\bibfnamefont {B.}~\bibnamefont
  {Schubert}}, \bibinfo {author} {\bibfnamefont {R.}~\bibnamefont
  {Maddamsetti}}, \bibinfo {author} {\bibfnamefont {J.}~\bibnamefont {Nyman}},
  \bibinfo {author} {\bibfnamefont {M.~R.}\ \bibnamefont {Farhat}}, \ and\
  \bibinfo {author} {\bibfnamefont {D.~S.}\ \bibnamefont {Marks}},\ }\href
  {\doibase 10.1038/s41564-018-0309-1} {\bibfield  {journal} {\bibinfo
  {journal} {Nat. Microbiol.}\ }\textbf {\bibinfo {volume} {4}},\ \bibinfo
  {pages} {328} (\bibinfo {year} {2019})}\BibitemShut {NoStop}%
\bibitem [{\citenamefont {Puranen}\ \emph {et~al.}(2018)\citenamefont
  {Puranen}, \citenamefont {Pesonen}, \citenamefont {Pensar}, \citenamefont
  {Xu}, \citenamefont {Lees}, \citenamefont {Bentley}, \citenamefont
  {Croucher},\ and\ \citenamefont {Corander}}]{Puranen-2017a}%
  \BibitemOpen
  \bibfield  {author} {\bibinfo {author} {\bibfnamefont {S.}~\bibnamefont
  {Puranen}}, \bibinfo {author} {\bibfnamefont {M.}~\bibnamefont {Pesonen}},
  \bibinfo {author} {\bibfnamefont {J.}~\bibnamefont {Pensar}}, \bibinfo
  {author} {\bibfnamefont {Y.~Y.}\ \bibnamefont {Xu}}, \bibinfo {author}
  {\bibfnamefont {J.~A.}\ \bibnamefont {Lees}}, \bibinfo {author}
  {\bibfnamefont {S.~D.}\ \bibnamefont {Bentley}}, \bibinfo {author}
  {\bibfnamefont {N.~J.}\ \bibnamefont {Croucher}}, \ and\ \bibinfo {author}
  {\bibfnamefont {J.}~\bibnamefont {Corander}},\ }\href@noop {} {\bibfield
  {journal} {\bibinfo  {journal} {Microb. Genom.}\ } (\bibinfo {year}
  {2018})}\BibitemShut {NoStop}%
\bibitem [{\citenamefont {Gao}\ \emph {et~al.}(2018)\citenamefont {Gao},
  \citenamefont {Zhou},\ and\ \citenamefont {Aurell}}]{GaoZhouAurell2018}%
  \BibitemOpen
  \bibfield  {author} {\bibinfo {author} {\bibfnamefont {C.-Y.}\ \bibnamefont
  {Gao}}, \bibinfo {author} {\bibfnamefont {H.-J.}\ \bibnamefont {Zhou}}, \
  and\ \bibinfo {author} {\bibfnamefont {E.}~\bibnamefont {Aurell}},\
  }\href@noop {} {\bibfield  {journal} {\bibinfo  {journal} {Phys. Rev. E}\
  }\textbf {\bibinfo {volume} {98}},\ \bibinfo {pages} {032407} (\bibinfo
  {year} {2018})}\BibitemShut {NoStop}%
\bibitem [{\citenamefont {Pensar}\ \emph {et~al.}(2019)\citenamefont {Pensar},
  \citenamefont {Puranen}, \citenamefont {Arnold}, \citenamefont {MacAlasdair},
  \citenamefont {Kuronen}, \citenamefont {Tonkin-Hill}, \citenamefont
  {Pesonen}, \citenamefont {Xu}, \citenamefont {Sipola}, \citenamefont
  {Sánchez-Busó}, \citenamefont {Lees}, \citenamefont {Chewapreecha},
  \citenamefont {Bentley}, \citenamefont {Harris}, \citenamefont {Parkhill},
  \citenamefont {Croucher},\ and\ \citenamefont {Corander}}]{Pensar2019}%
  \BibitemOpen
  \bibfield  {author} {\bibinfo {author} {\bibfnamefont {J.}~\bibnamefont
  {Pensar}}, \bibinfo {author} {\bibfnamefont {S.}~\bibnamefont {Puranen}},
  \bibinfo {author} {\bibfnamefont {B.}~\bibnamefont {Arnold}}, \bibinfo
  {author} {\bibfnamefont {N.}~\bibnamefont {MacAlasdair}}, \bibinfo {author}
  {\bibfnamefont {J.}~\bibnamefont {Kuronen}}, \bibinfo {author} {\bibfnamefont
  {G.}~\bibnamefont {Tonkin-Hill}}, \bibinfo {author} {\bibfnamefont
  {M.}~\bibnamefont {Pesonen}}, \bibinfo {author} {\bibfnamefont
  {Y.}~\bibnamefont {Xu}}, \bibinfo {author} {\bibfnamefont {A.}~\bibnamefont
  {Sipola}}, \bibinfo {author} {\bibfnamefont {L.}~\bibnamefont
  {Sánchez-Busó}}, \bibinfo {author} {\bibfnamefont {J.~A.}\ \bibnamefont
  {Lees}}, \bibinfo {author} {\bibfnamefont {C.}~\bibnamefont {Chewapreecha}},
  \bibinfo {author} {\bibfnamefont {S.~D.}\ \bibnamefont {Bentley}}, \bibinfo
  {author} {\bibfnamefont {S.~R.}\ \bibnamefont {Harris}}, \bibinfo {author}
  {\bibfnamefont {J.}~\bibnamefont {Parkhill}}, \bibinfo {author}
  {\bibfnamefont {N.~J.}\ \bibnamefont {Croucher}}, \ and\ \bibinfo {author}
  {\bibfnamefont {J.}~\bibnamefont {Corander}},\ }\href@noop {} {\bibfield
  {journal} {\bibinfo  {journal} {Nucleic Acids Res.}\ }\textbf {\bibinfo
  {volume} {47}},\ \bibinfo {pages} {e112} (\bibinfo {year}
  {2019})}\BibitemShut {NoStop}%
\bibitem [{\citenamefont {Gao}\ \emph {et~al.}(2019)\citenamefont {Gao},
  \citenamefont {Cecconi}, \citenamefont {Vulpiani}, \citenamefont {Zhou},\
  and\ \citenamefont {Aurell}}]{Gao2019}%
  \BibitemOpen
  \bibfield  {author} {\bibinfo {author} {\bibfnamefont {C.-Y.}\ \bibnamefont
  {Gao}}, \bibinfo {author} {\bibfnamefont {F.}~\bibnamefont {Cecconi}},
  \bibinfo {author} {\bibfnamefont {A.}~\bibnamefont {Vulpiani}}, \bibinfo
  {author} {\bibfnamefont {H.-J.}\ \bibnamefont {Zhou}}, \ and\ \bibinfo
  {author} {\bibfnamefont {E.}~\bibnamefont {Aurell}},\ }\href {\doibase
  10.1088/1478-3975/aafbe0} {\bibfield  {journal} {\bibinfo  {journal} {Phys.
  Biol.}\ }\textbf {\bibinfo {volume} {16}},\ \bibinfo {pages} {026002}
  (\bibinfo {year} {2019})}\BibitemShut {NoStop}%
\bibitem [{\citenamefont {Kimura}(1956)}]{Kimura1956}%
  \BibitemOpen
  \bibfield  {author} {\bibinfo {author} {\bibfnamefont {M.}~\bibnamefont
  {Kimura}},\ }\href@noop {} {\bibfield  {journal} {\bibinfo  {journal}
  {Evolution}\ }\textbf {\bibinfo {volume} {10}},\ \bibinfo {pages} {278}
  (\bibinfo {year} {1956})}\BibitemShut {NoStop}%
\bibitem [{\citenamefont {Kimura}(1964)}]{Kimura1964}%
  \BibitemOpen
  \bibfield  {author} {\bibinfo {author} {\bibfnamefont {M.}~\bibnamefont
  {Kimura}},\ }\href {\doibase 10.2307/3211856} {\bibfield  {journal} {\bibinfo
   {journal} {J. Appl. Probab.}\ }\textbf {\bibinfo {volume} {1}},\ \bibinfo
  {pages} {177–232} (\bibinfo {year} {1964})}\BibitemShut {NoStop}%
\bibitem [{\citenamefont {Kimura}(1965)}]{Kimura1965}%
  \BibitemOpen
  \bibfield  {author} {\bibinfo {author} {\bibfnamefont {M.}~\bibnamefont
  {Kimura}},\ }\href@noop {} {\bibfield  {journal} {\bibinfo  {journal}
  {Genetics}\ }\textbf {\bibinfo {volume} {52}},\ \bibinfo {pages} {875}
  (\bibinfo {year} {1965})}\BibitemShut {NoStop}%
\bibitem [{\citenamefont {Neher}\ and\ \citenamefont
  {Shraiman}(2009)}]{NeherShraiman2009}%
  \BibitemOpen
  \bibfield  {author} {\bibinfo {author} {\bibfnamefont {R.~A.}\ \bibnamefont
  {Neher}}\ and\ \bibinfo {author} {\bibfnamefont {B.~I.}\ \bibnamefont
  {Shraiman}},\ }\href {\doibase 10.1073/pnas.0812560106} {\bibfield  {journal}
  {\bibinfo  {journal} {Proc. Natl. Acad. Sci.}\ }\textbf {\bibinfo {volume}
  {106}},\ \bibinfo {pages} {6866} (\bibinfo {year} {2009})}\BibitemShut
  {NoStop}%
\bibitem [{\citenamefont {Neher}\ and\ \citenamefont
  {Shraiman}(2011)}]{NeherShraiman2011}%
  \BibitemOpen
  \bibfield  {author} {\bibinfo {author} {\bibfnamefont {R.~A.}\ \bibnamefont
  {Neher}}\ and\ \bibinfo {author} {\bibfnamefont {B.~I.}\ \bibnamefont
  {Shraiman}},\ }\href {\doibase 10.1103/RevModPhys.83.1283} {\bibfield
  {journal} {\bibinfo  {journal} {Rev. Mod. Phys.}\ }\textbf {\bibinfo {volume}
  {83}},\ \bibinfo {pages} {1283} (\bibinfo {year} {2011})}\BibitemShut
  {NoStop}%
\bibitem [{\citenamefont {Neher}\ and\ \citenamefont
  {Zanini}(2012)}]{FFPopSim}%
  \BibitemOpen
  \bibfield  {author} {\bibinfo {author} {\bibfnamefont {R.}~\bibnamefont
  {Neher}}\ and\ \bibinfo {author} {\bibfnamefont {F.}~\bibnamefont {Zanini}},\
  }\href@noop {} {\enquote {\bibinfo {title} {Ffpopsim},}\ }\bibinfo
  {howpublished} {http://code.google.com/p/ffpopsim/} (\bibinfo {year}
  {2012})\BibitemShut {NoStop}%
\bibitem [{\citenamefont {Fisher}(1930)}]{Fisher-book}%
  \BibitemOpen
  \bibfield  {author} {\bibinfo {author} {\bibfnamefont {R.}~\bibnamefont
  {Fisher}},\ }\href@noop {} {\emph {\bibinfo {title} {The Genetical Theory of
  Natural Selection}}}\ (\bibinfo  {publisher} {Clarendon},\ \bibinfo {year}
  {1930})\BibitemShut {NoStop}%
\bibitem [{\citenamefont {Blythe}\ and\ \citenamefont
  {McKane}(2007)}]{Blythe2007}%
  \BibitemOpen
  \bibfield  {author} {\bibinfo {author} {\bibfnamefont {R.~A.}\ \bibnamefont
  {Blythe}}\ and\ \bibinfo {author} {\bibfnamefont {A.~J.}\ \bibnamefont
  {McKane}},\ }\href {\doibase 10.1088/1742-5468/2007/07/p07018} {\bibfield
  {journal} {\bibinfo  {journal} {J. Stat. Mech.: Theory Exp.}\ }\textbf
  {\bibinfo {volume} {2007}},\ \bibinfo {pages} {P07018} (\bibinfo {year}
  {2007})}\BibitemShut {NoStop}%
\bibitem [{\citenamefont {Ekeberg}\ \emph {et~al.}(2014)\citenamefont
  {Ekeberg}, \citenamefont {Hartonen},\ and\ \citenamefont
  {Aurell}}]{Ekeberg-2014a}%
  \BibitemOpen
  \bibfield  {author} {\bibinfo {author} {\bibfnamefont {M.}~\bibnamefont
  {Ekeberg}}, \bibinfo {author} {\bibfnamefont {T.}~\bibnamefont {Hartonen}}, \
  and\ \bibinfo {author} {\bibfnamefont {E.}~\bibnamefont {Aurell}},\ }\href
  {\doibase 10.1016/j.jcp.2014.07.024} {\bibfield  {journal} {\bibinfo
  {journal} {J. Comput. Phys.}\ }\textbf {\bibinfo {volume} {276}},\ \bibinfo
  {pages} {341} (\bibinfo {year} {2014})}\BibitemShut {NoStop}%
\bibitem [{\citenamefont {Gao}(2018)}]{Gao-github}%
  \BibitemOpen
  \bibfield  {author} {\bibinfo {author} {\bibfnamefont {C.-Y.}\ \bibnamefont
  {Gao}},\ }\href {github.com/gaochenyi/CC-PLM} {\enquote {\bibinfo {title}
  {gaochenyi/cc-plm},}\ }\bibinfo {howpublished} {Github} (\bibinfo {year}
  {2018})\BibitemShut {NoStop}%
\end{thebibliography}%

\end{document}